\documentclass[preprint,aps,floats,showpacs,superscriptaddress,nofootinbib]{revtex4} 
\usepackage{graphicx} 
\usepackage{epsfig} 
\usepackage{amsmath} 
 
\def\be{\begin{equation}} 
\def\ee{\end{equation}} 
\def\beq{\begin{equation}} 
\def\eeq{\end{equation}} 
\def\bea{\begin{eqnarray}} 
\def\eea{\end{eqnarray}} 
\def\dmsq{\Delta m^2}
\def\barr{\begin{eqnarray}}
\def\earr{\end{eqnarray}}
\def\mtilde{\widetilde{M}}

\newcommand{\lsim}{\mbox{\raisebox{-.6ex}{~$\stackrel{<}{\sim}$~}}}
\newcommand{\gsim}{\mbox{\raisebox{-.6ex}{~$\stackrel{>}{\sim}$~}}} 
\begin{document} 
 
\rightline{TIFR/TH/08-54}

\title{\bf Texture zeroes and discrete flavor symmetries in light and
heavy Majorana neutrino mass matrices: a bottom-up approach} 

\author{Amol Dighe} 
\email{amol@theory.tifr.res.in} 
\affiliation{Tata Institute of 
Fundamental Research, Homi Bhabha Road, Mumbai 400005, INDIA} 

\author{Narendra Sahu} 
\email{n.sahu@lancaster.ac.uk} 
\affiliation{Department of Physics, Lancaster University, 
Lancaster, LA1 4YB, UK}

\begin{abstract}
 
Texture zeroes in neutrino mass matrix $M_\nu$ 
may give us hints about the symmetries involved in
neutrino mass generation.
We examine the viability of such texture zeroes in a model independent
way through a bottom-up approach.
Using constraints from the neutrino oscillation data,
we develop an analytic framework that can identify these
symmetries and quantify deviations from them.
We analyze the textures of $M_\nu$ as well as those of $M_M$,
the mass matrix of heavy Majorana neutrinos in the
context of Type-I seesaw.
We point out how the viability of textures depends on the 
absolute neutrino mass scale, the neutrino mass ordering
and the mixing angle $\theta_{13}$.
We also examine the compatibility of discrete flavor symmetries 
like $\mu$--$\tau$ exchange and $S_3$ permutation  with the current data.
We show that the $\mu-\tau$ exchange symmetry for $M_\nu$ can be
satisfied for any value of the absolute neutrino mass,
but for $M_\nu$ to satisfy the $S_3$ symmetry, neutrino masses have 
to be quasi-degenerate. 
On the other hand, both these symmetries are currently
allowed for $M_M$ for all values of absolute neutrino mass 
and both mass orderings.

\end{abstract} 
\maketitle 

\section{Introduction} 

The current low energy neutrino oscillation data
\cite{nu-fits} 
indicate that all three of the physical left-handed neutrinos have 
different masses and they mix among themselves. 
If the neutrinos are Majorana, the neutrino mass matrix $M_\nu$ 
in the flavor basis is diagonalized by the unitary 
Pontecorvo-Maki-Nakagawa-Sakata matrix $U_{\rm PMNS}$ 
\cite{pontecorvo,mns}
through
\beq
M_\nu^{\rm diag} = U_{\rm PMNS}^\dagger M_\nu U_{\rm PMNS}^* 
\; ,
\qquad {\rm i.e.} \qquad
M_\nu = U_{\rm PMNS} M_\nu^{\rm diag} U_{\rm PMNS}^T 
\; .
\label{u-m-ut}
\eeq
For Majorana neutrinos, $U_{\rm PMNS}$ is given by
\begin{eqnarray} 
U_{\rm PMNS}= 
U_\chi \cdot
\left( 
\begin{array}{ccccc} 
c_{12}c_{13} & & s_{12}c_{13} & &s_{13}e^{-i\delta} \\ 
-s_{12}c_{23}-c_{12}s_{23}s_{13}e^{i\delta} & & c_{12}c_{23}- 
s_{12}s_{23}s_{13}e^{i\delta} & & s_{23}c_{13} \\ 
s_{12}s_{23}-c_{12}c_{23}s_{13}e^{i\delta} & & -c_{12}s_{23}-s_{12} 
c_{23}s_{13}e^{i\delta} & & c_{23}c_{13} 
\end{array} 
\right)\, \cdot \,U_{\phi} \; ,
\label{mns-matrix} 
\end{eqnarray} 
where $c_{ij}$ and $s_{ij}$ stand for $\cos \theta_{ij}$ and
$\sin \theta_{ij}$ respectively. Here $U_{\phi}={\rm diag}(e^{i\phi_1}, 
e^{i\phi_2},1)$, with the Majorana phases $\phi_1$ and $\phi_2$ 
defined in such a way that the diagonal elements of 
$M_\nu^{\rm diag}$ are given by 
\beq 
M_\nu^{\rm diag}={\rm diag}(m_1, m_2, m_3)\, .
\label{absolute_masses}
\eeq
Here $m_i$ ($i=1,2,3$) correspond to the neutrino masses,
which are chosen to be real and positive. 
The Dirac phase $\delta$ accounts for the charge parity (CP) 
violation in the lepton number conserving processes. 
The phase matrix $U_\chi \equiv {\rm diag}(e^{i\chi_e}, e^{i\chi_\mu},
e^{i\chi_\tau})$ consists of the three ``flavor phases''\footnote{
These phases are often referred to in the literature as ``unphysical phases''.
Though these phases have no relevance for the low energy neutrino
phenomenology and cannot be determined through low energy
measurements, their values may be predictable
within the context of specific models with new physics at the high scale.}
$\chi_\alpha$ that correspond to the multiplication of a neutrino
flavor eigenstate $\nu_\alpha$ by $e^{i\chi_\alpha}$. 
Note that once the mixing angles $\theta_{ij}$ have been defined
to be in the first quadrant, the Dirac phase $\delta$ can take
values in $[0,2\pi)$ and the phases $\chi_\alpha, \phi_i$ can take
values between $[0,\pi)$. All the angles and phases are then
uniquely defined.

A global analysis of the current neutrino oscillation data at 
$3\sigma$ C.L. yields~\cite{nu-fits} 
\beq 
0.25 < \sin^2\theta_{12} < 0.37 \; , \qquad 
0.36 < \sin^2\theta_{23} < 0.67 \;,  \qquad
\sin^2 \theta_{13} < 0.056 \;.
\eeq
While the absolute mass scale of the neutrinos is not yet fixed, 
the two mass-squared differences 
have already been determined to a good degree of accuracy: 
\begin{eqnarray} 
\Delta m^2_{\odot} &\equiv& m_2^2 - m_1^2 = 
(7.06 \cdots 8.34)\times 10^{-5}~ {\rm eV}^2 \; ,
\nonumber\\ 
\Delta m^2_{\rm atm} &\equiv& m_3^2 - 
\left(\frac{m_1 + m_2}{2} \right)^2 =  \pm (2.07 \cdots 2.75) 
\times 10^{-3}~ {\rm eV}^2\; .
\label{mass-sq-diff} 
\end{eqnarray} 
It is not known whether the neutrino mass ordering is normal ($m_1 < m_2 < m_3$)
or inverted ($m_3 < m_1 < m_2$).
The Dirac phase $\delta$ and Majorana phases $\phi_{1,2}$ are completely 
unknown.
The absolute values of neutrino masses cannot be probed by oscillation 
experiments, the direct limit on the neutrino mass scale $m_0$
is obtained by the tritium beta decay experiments \cite{pdg} as  
$m_0 < 2.2$ eV. The most stringent 
constraint on $m_0$ however comes from cosmology: the WMAP data
implies \cite{wmap}
\beq
\sum m_i \lsim 1 ~{\rm eV} \; .
\label{sum-mi}
\eeq
In this paper, we shall take the upper bound on each neutrino mass 
conservatively to be $m_i < 0.5$ eV.

Given the absolute values of the neutrino masses and the complete 
matrix $U_{\rm PMNS}$, the neutrino mass matrix $M_\nu$ in the flavor 
basis can be reconstructed through Eq.~(\ref{u-m-ut}).
The structure of this matrix may reveal the presence of flavor 
symmetries in the neutrino sector. 
In this paper, we consider multiple texture zeroes of $M_\nu$
\cite{one-zero,texture-zero,two-zero-textures} as well as symmetries like 
the $\mu-\tau$ exchange~\cite{mu-tau} and $S_3$ permutation \cite{s3}, 
which predict certain relations between the elements of $M_\nu$.

The symmetry-based relations among the elements of $M_\nu$ and 
texture zeroes of $M_\nu$ have been 
explored earlier mainly by adopting a top-down approach~\cite{top-down}. 
In this approach an appropriate symmetry is imposed on the 
neutrino mass matrix, which in turn gives a prediction for 
the neutrino mixing parameters that can then be checked against the 
available data. 
We take the bottom-up approach, starting with our current knowledge 
about neutrino masses and mixings, and checking if a certain texture 
zero combination or a symmetry-based relation is allowed. 
This allows us to test in a model independent manner the symmetries 
present in neutrino mass generation mechanisms.
It also enables us to determine which future measurements can
act as tests of these symmetries.

In the present approach the elements of $M_\nu$ are expressed as 
functions of the absolute neutrino masses, the mixing angles as well as 
the Dirac, Majorana and flavor phases. 
Our current complete ignorance about these phases allows a lot of freedom for 
the elements of $M_\nu$ in spite of the relatively well measured values 
of the masses and mixing angles. 
Even with this freedom, some of the texture zero combinations and 
symmetries are clearly forbidden, as has been numerically 
verified~\cite{forbidden-textures}. 
We develop an analytical treatment, using perturbative expansion in 
appropriate small parameters, and demonstrate the analytical rationale 
behind the ruling out of some of these relations. 
This also leads us to the result that the additional knowledge of 
the absolute mass scale of the neutrinos and the mixing angle $\theta_{13}$
will be crucial in testing for these relations in  near future.

The seesaw mechanism~\cite{seesaw} is one of the most favored and 
explored mechanisms for neutrino mass generation, which gives rise to 
light Majorana neutrinos that can satisfy the low 
energy neutrino oscillation data, as well as to heavy Majorana neutrinos 
that may play an important role in leptogenesis~\cite{leptogenesis}. 
If the neutrino masses are generated from a Type-I seesaw mechanism 
where three singlet heavy Majorana neutrinos are added to the Standard 
Model (SM), then we have the effective neutrino mass matrix
\begin{equation}
M_\nu = - m_D M_M^{-1} (m_D)^T\,,
\label{typeI-seesaw}
\end{equation}
where $m_D$ is the Dirac mass matrix of neutrinos,
and $M_M$ is the Majorana mass matrix for the right-handed heavy Majorana 
neutrinos. 
If the heavy Majorana neutrinos are written in a basis where 
$m_D$ is diagonal, the texture zeroes as well as symmetry relations 
between elements of $M_M$ can be related to those 
of the inverse neutrino matrix $M_\nu^{-1}$ in a straightforward 
manner~\cite{ma-determinant}. 
The same analytical treatment developed for $M_\nu$ can then be 
extended to test the symmetry relations for $M_M$ in this basis. 
We perform this analysis, with a particular emphasis on the dependence of 
these relations on the absolute masses of the light neutrinos.
 
The paper is organized as follows. In Sec.~\ref{formalism}, we introduce 
our formalism and set up the analytical framework under which the 
symmetry relations may be examined. In Sec.~\ref{mnu-zeroes}
and Sec.~\ref{mM-zeroes}, we test the texture zeroes of $M_\nu$
and $M_M$ respectively, numerically as well as analytically.
In \ref{mu-tau} and \ref{mM}, we examine the
$\mu-\tau$ exchange symmetry and $S_3$ permutation symmetry
for $M_\nu$ and $M_M$ respectively.
Sec.~\ref{concl} concludes.

\section{The analytical framework}
\label{formalism} 

\subsection{parameterization of neutrino masses and mixing}
\label{parametrization}

\begin{figure}[h!] 
\begin{center} 
\epsfig{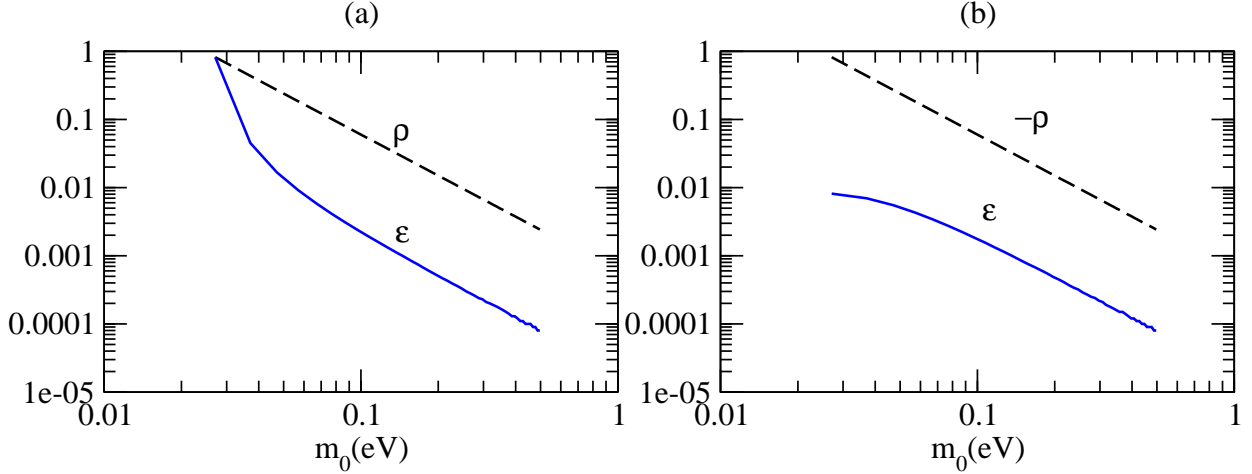} 
\caption{The parameters $\rho$ and $\epsilon$ as functions of $m_0$ 
for (a) normal ordering and (b) 
inverted ordering of neutrino masses,
for best-fit values of $\Delta m_{\rm atm}^2$ and 
$\Delta m_{\odot}^2$.}
\label{norinv} 
\end{center} 
\end{figure}

We parameterize the absolute values of neutrino masses in terms of three parameters 
$m_0$, $\epsilon$ and $\rho$ as~\cite{dgr2} 
\beq
m_1 = m_0 (1-\rho)(1-\epsilon) \; , \qquad 
m_2 = m_0 (1-\rho)(1+\epsilon) \; , \qquad 
m_3 = m_0 (1+\rho) \; ,
\label{abs_neu_mass} 
\eeq
where $m_0$ sets the overall mass scale of neutrinos, while  
the dimensionless parameters $\rho$ and $\epsilon$ can be expressed  
in terms of the solar and atmospheric mass scales as 
\beq 
\rho = \frac{\Delta m_{\rm atm}^2}{4 m_0^2} \; ,  \qquad 
\epsilon = \frac{\Delta m_\odot^2}{4 m_0^2 (1-\rho)^2} \; . 
\label{rhoeps-value} 
\eeq
Clearly, $\rho$ is positive (negative) for normal (inverted)
mass ordering of neutrinos. The sum of neutrino masses may be expressed 
in terms of the above parameters as
\beq
\sum_i m_i = 3 m_0\left( 1-\frac{\rho}{3} \right)\lsim 1~{\rm eV} \; .
\eeq
The condition $0 < m_i < 0.5$ eV then yields  
\beq
m_0 \gtrsim 0.025 ~ {\rm eV} \; , \qquad
 2.43 \times 10^{-3} < |\rho| < 1 \; , \qquad
 8 \times 10^{-5} < \epsilon < 1 \; .
\eeq
The value of $|\rho|$ approaches unity as $m_0$ approaches its lowest allowed value.
The value of $\epsilon$ can be $> 0.01$ only for normal mass ordering and $m_0 < 0.06$ eV, 
whereas $\epsilon \ll |\rho|$ everywhere except for $m_0 \approx 0.025$ eV. Taking the best-fit 
values of solar and atmospheric neutrino masses, in Fig.~\ref{norinv} we show the values of 
$\rho$ and $\epsilon$ as functions of $m_0$ for normal as well as inverted hierarchies.
For the purpose of this paper, we divide the neutrino parameter space into
three scenarios:\\
(i) Normal mass ordering with hierarchical masses (NH), where 
$m_1 \ll m_2 \ll m_3$. 
The current data give
$\rho \approx 0.85$ and $\epsilon \approx 0.92$ in the extreme
limit, however these values decrease rather rapidly as $m_0$
increases, as can be seen from Fig.~\ref{norinv}.
This scenario can then be analyzed through a 
perturbative expansion in the set of the small parameters  
$\tilde\rho \equiv 1 - \rho$ and $\tilde\epsilon \equiv 1 - \epsilon$
in the extreme limit, however one has to be careful while treating 
quantities like $(1-\rho^2)$, which stays higher than 0.3. \\
(ii) Inverted mass ordering with hierarchical masses (IH), such that
$m_3 \ll m_1 < m_2$. In this case, 
$\epsilon \ll 1$ and $|1 + \rho| \ll 1$, so we can use
a perturbative expansion in the small parameters 
$\widehat\rho \equiv 1 + \rho$ and $\epsilon$. \\
(iii) Quasidegenerate neutrinos (QD), where $m_1 \simeq m_2 \simeq m_3$,
with either mass ordering. In this case, $|\rho|,\epsilon \ll 1$, 
so that we can use these two quantities as small parameters.

In addition, at appropriate places we shall also consider $\theta_{13}$
and $\tilde{\theta}_{23} \equiv \theta_{23} - \pi/4$ as small 
parameters in order to facilitate a perturbative expansion.

\subsection{Elements in $M_\nu$, $M_\nu^{-1}$,
$\widetilde{M}_\nu$ and $M_M$}
\label{expansions}

Let the low energy neutrino mass matrix in the flavor basis 
be written as

\beq
M_\nu= \left(
\begin{array}{ccc} 
a & b & c\\ 
b & d & e \\ 
c & e & f\end{array} \right)\; . 
\label{mnu-def}
\end{equation} 
Since the neutrinos are Majorana, $M_\nu$ is symmetric.
In terms of the parameterization of neutrino masses in 
Sec.~\ref{parametrization}, mixing angles and CP violating
phases, the elements of $M_\nu$ may be written as
\barr
a & = & m_0 e^{2i\chi_e}
~\bigg[ 
e^{2 i\phi_1} (-1+\rho)(-1+\epsilon) c_{12}^2 c_{13}^2\nonumber\\ 
& & -e^{2i\phi_2}(-1+\rho)(1+\epsilon) c_{13}^2 s_{12}^2
+ e^{-2i\delta}(1+\rho) s_{13}^2 \bigg] \; ,
\nonumber \\
b & = & m_0 e^{i(\chi_e+\chi_\mu)} c_{13}
~\bigg[ e^{-i\delta}(1+\rho) s_{23} s_{13} \nonumber \\ 
& & - e^{2i\phi_1}(-1+\rho)(-1+\epsilon) c_{12} \left( c_{23} s_{12} 
+e^{i\delta} c_{12} s_{23} s_{13} \right) \nonumber \\
& & + e^{2i \phi_2}(-1+\rho)(1+\epsilon) s_{12} 
\left( - c_{12} c_{23} + e^{i\delta} s_{12} s_{23} s_{13} \right) 
\bigg] \; ,
\nonumber \\
c & = & m_0 e^{i(\chi_e+\chi_\tau)} c_{13}
~\bigg[ e^{-i\delta} (1+\rho) c_{23} s_{13} \nonumber \\
& & - e^{2i \phi_1}(-1+\rho)(-1+\epsilon) c_{12} 
\left( -s_{12} s_{23} + e^{i\delta} c_{12} c_{23} s_{13} \right) \nonumber \\
& & + e^{2i\phi_2}(-1+\rho)(1+\epsilon) s_{12}\left( c_{12} s_{23} + 
e^{i\delta} c_{23} s_{12} s_{13} \right) \bigg]  \; ,
\nonumber \\
d & = & m_0 e^{2 i \chi_\mu} 
~\bigg[ 
(1+\rho) c_{13}^2 s_{23}^2 \nonumber \\
& & + e^{2i \phi_1}(-1+\rho)
(-1+\epsilon) \left( c_{23} s_{12}+e^{i\delta} c_{12} s_{23} s_{13} \right)^2 \nonumber \\
& & -e^{2i \phi_2}(-1+\rho) (1+\epsilon) 
\left( c_{12} c_{23} -e^{i\delta} s_{12} s_{23} s_{13} \right)^2 
\bigg]  \; , 
\nonumber \\
e & = & m_0 e^{i(\chi_\mu+\chi_\tau)}
~\bigg[ 
(1+\rho) c_{23} c_{13}^2 s_{23} \nonumber \\ 
& & -e^{2i\phi_1}(-1+\rho)(-1+\epsilon) 
\left( s_{12} s_{23}- e^{i\delta} c_{12} c_{23} s_{13} \right) 
\left( c_{23} s_{12}+ e^{i\delta} c_{12} s_{23} s_{13} \right)
\nonumber \\
& & +e^{2i\phi_2}(-1+\rho)(1+\epsilon) 
\left( c_{12} s_{23} + e^{i\delta} c_{23} s_{12} s_{13}\right) 
\left(c_{12} c_{23}-e^{i\delta} s_{12} s_{23} s_{13} \right) \bigg] \; ,
\nonumber \\
f & = & m_0  e^{2i\chi_\tau} 
~\bigg[ 
(1+\rho) c_{23}^2 c_{13}^2 \nonumber \\
& & +e^{2i\phi_1}(-1+\rho)
(-1+\epsilon) \left( s_{12} s_{23}
-e^{i\delta} c_{12} c_{23} s_{13}\right)^2 
\nonumber \\
& & -e^{2i \phi_2}(-1+\rho)(1+\epsilon)\left( c_{12} s_{23} 
+ e^{i\delta} c_{23} s_{12} s_{13}\right)^2 \bigg] \; .
\label{abcdef-def}
\earr

The inverse of the neutrino mass matrix, $M_\nu^{-1}$, can be written as
\beq
M_\nu^{-1} = \frac{\widetilde{M}}{Det(M_\nu)} \equiv
\frac{1}{Det(M_\nu)} 
\begin{pmatrix} 
A & B & C \\ 
B & D & E \\ 
C & E & F \end{pmatrix}\; ,
\label{mnu-inv-def}
\eeq
where 
\beq
Det(M_\nu) = m_0^3 (1-\rho)^2 (1+\rho) (1-\epsilon^2) e^{2i\sum\chi} \; 
\eeq
is the determinant of $M_\nu$,
with $\sum\chi\equiv \chi_e+\chi_\mu+\chi_\tau$.
Here $\widetilde{M}$ is the adjoint neutrino mass matrix.
From Eq.~(\ref{mnu-inv-def}) it is obvious that texture zeroes 
in $M_\nu^{-1}$ are the same as those in $\widetilde{M}$. 
The elements of $\widetilde{M}$ can be written in terms of the masses 
and the elements of $U \equiv U_{\rm PMNS}$ matrix as
\barr
A &=& 
m_0^2 e^{2i(\chi_\mu+\chi_\tau)} \bigl[ (1-\rho)^2(1-\epsilon^2)\left( U_{21}U_{32}-U_{22}U_{31} 
\right)^2 \nonumber\\
&& + (1-\rho^2) (1-\epsilon) \left(U_{21} U_{33}-U_{23}U_{31} \right)^2 
 +  (1-\rho^2)(1+\epsilon)\left( U_{22} U_{33}-U_{23} U_{32}\right)^2 \bigr] 
\nonumber\\
B &=& 
m_0^2 e^{i(\chi_e + \chi_\mu + 2 \chi_\tau)}\bigl[ (1-\rho)^2(1-\epsilon^2) \left( U_{21}U_{32} - 
U_{22}U_{31} \right) \left( U_{31}U_{12}-U_{11}U_{32} \right) \nonumber \\
& & + (1-\rho^2) (1-\epsilon) \left( U_{31} U_{13} -U_{33}U_{11} \right) 
\left( U_{21}U_{33}-U_{23}U_{31} \right) \nonumber \\
& & + (1-\rho^2)(1+\epsilon) \left( U_{22} U_{33}-U_{23}U_{32} \right) 
\left( U_{32}U_{13}-U_{12}U_{33} \right) \bigr] \nonumber\\
C &=& 
m_0^2 e^{i(\chi_e + 2\chi_\mu + \chi_\tau)} \bigl[ (1-\rho)^2(1-\epsilon^2) \left( U_{21}U_{32}-U_{22}U_{31} \right) 
\left( U_{22}U_{11}-U_{12}U_{21} \right) \nonumber \\
& & + (1-\rho^2) (1-\epsilon) \left( U_{11}U_{23}-U_{13}U_{21} \right) 
\left( U_{21}U_{33}-U_{31}U_{23} \right) \nonumber \\
& & + (1-\rho^2)(1+\epsilon) \left( U_{12}U_{23}-U_{13}U_{22} \right) 
\left( U_{22}U_{33}-U_{32}U_{23} \right) \bigr]\nonumber\\
D &=& 
m_0^2 e^{2i(\chi_e + \chi_\tau)}\bigl[ (1-\rho)^2(1-\epsilon^2) \left( U_{11}U_{32}-U_{12}U_{31} \right)^2+ 
(1-\rho^2) (1-\epsilon) \left( U_{11}U_{33}-U_{13}U_{31} \right)^2 \nonumber\\ 
&& + (1-\rho^2)(1+\epsilon) \left( U_{12}U_{33}-U_{32}U_{13} \right)^2 \bigr] 
\nonumber\\
E &=& 
m_0^2 e^{i(2 \chi_e + \chi_\mu + \chi_\tau)}\bigl[ (1-\rho)^2(1-\epsilon^2) \left( U_{11}U_{32}-U_{12}U_{31} \right) 
\left( U_{12}U_{21}-U_{11}U_{22} \right) \nonumber \\
& & + (1-\rho^2) (1-\epsilon) \left( U_{11}U_{33}-U_{13}U_{31} \right) 
\left( U_{21}U_{13}-U_{11}U_{23} \right) \nonumber \\
& & + (1-\rho^2)(1+\epsilon) \left( U_{12}U_{33}-U_{13}U_{32} \right) 
\left( U_{22}U_{13}-U_{12}U_{23} \right) \bigr] \nonumber\\
F &=& 
m_0^2 e^{2 i(\chi_e + \chi_\mu)}\bigl[ (1-\rho)^2(1-\epsilon^2) \left( U_{11}U_{22}-U_{12}U_{21} \right)^2
+ (1-\rho^2) (1-\epsilon) \left( U_{11}U_{23}-U_{13}U_{21} \right)^2 \nonumber\\ 
&& + (1-\rho^2)(1+\epsilon) \left( U_{12}U_{23}-U_{13}U_{22} \right)^2\bigr]\,.
\label{mtilde-elements}
\earr

In order to analyze the heavy majorana neutrino mas matrix $M_M$,
we invert Eq.~(\ref{typeI-seesaw}) to obtain
\beq
M_M= - m_D^T M_\nu^{-1} m_D\,.
\eeq
Following \cite{ma-determinant}, 
we choose the ``flavor'' basis for heavy Majorana neutrinos in which
the Dirac mass matrix is real and diagonal,
\beq
m_D = {\rm diag}(x,y,z) \; .
\label{md-def}
\eeq
In this basis, $M_M$ may be written as
\beq
M_M= - \frac{1}{Det(M_\nu)} 
\begin{pmatrix} 
x^2 A & x y B & x z C \\ 
x y B & y^2 D & y z E \\ 
x z C & y z E & z^2 F \end{pmatrix}\; . 
\label{mM-def}
\eeq
Again, the texture zeroes of $M_M$ are same as the texture
zeroes of $M_\nu^{-1}$, and hence those of $\widetilde{M}$,
which have relatively tractable analytical expressions.
The discrete symmetries like $\mu-\tau$ exchange or $S_3$,
on the other hand, also depend on the values of the Dirac
masses. However even in that case, we can test for certain relations
between elements of $M_M$ that are independent of these
Dirac masses, as we shall see in Sec.~\ref{mM}.

\subsection{Quantifying deviation from exact symmetry
in the bottom-up approach}
\label{whatiszero}

In the traditional top-down approach, discrete flavor symmetries like 
$\mu-\tau$ exchange, $S_3$-permutation, etc. 
are assumed in the neutrino mass matrix, 
which predict the mixing parameters measured in the low energy neutrino 
oscillation data. 
The main purpose of these symmetries is to understand why the (1-3) family
mixing is small, while the (2-3) family mixing is almost maximal and 
the (1-2) family mixing is large but not maximal. 
Traditionally these symmetries are employed to set $U_{13}=0$. 
The dynamical breaking of these symmetries then may predict a non-zero 
$U_{13}$, which is then compared with the experimental value. 
Since the current low energy neutrino oscillation data have large 
uncertainties, the data allow an enormous freedom to propose such 
discrete flavor symmetries in the top-down approach. 

However it is also crucial to examine, by starting with the available low 
energy neutrino oscillation data, the parameter space of neutrino mass 
matrix where such discrete flavor symmetries can be realized. 
We call it the {\it bottom-up approach}. 
In this approach, since the low energy data have intrinsic uncertainties, 
the symmetry relations can only be said to be satisfied 
approximately. One therefore needs to quantify when one may declare 
the relevant symmetry to be allowed.

In $M_\nu$, the magnitudes of all the elements are expected to be 
$\sim m_0$, as can be seen 
from Eqs.~(\ref{abcdef-def}), taking into account
that the sine and cosine of $\theta_{12}, \theta_{23} \sim {\cal O}(1)$ 
and also assuming that $\phi_i \sim {\cal O}(1)$ in the absence of
any symmetry principle. 
If an element $|M_\nu(i,j)|$ is $\ll m_0$, it 
is either an accidental cancellation or the signature of a discrete 
symmetry at work. 
We take the position that for a sufficiently small value of $\xi$, 
the observation $|M_\nu(i,j)|/m_0 < \xi$ would indicate that the 
symmetry that would make $M_\nu(i,j)=0$ is present. 
We choose $\xi = 10^{-2}$, which is motivated by the accuracy to which 
the mixing angles are currently known. 
It also indicates the extent to which we tolerate the breaking of the 
discrete symmetries under consideration. 
In other words, when $|M_\nu(i,j)|/m_0 < 10^{-2}$, we consider $M_\nu(i,j)$ 
to be effectively zero.
Thus, we declare a texture zero viable if
\be
{\rm Min}\left( \frac{|M_\nu(i,j)|}{m_0} \right) < 10^{-2}\,,
\label{zero:M}
\ee 
where the minimization is over all the allowed values ($3\sigma$)
of the mixing parameters at a particular value of $m_0$.
If this condition is not satisfied, then the symmetry that would lead to
$M_\nu(i,j)=0$ is ruled out.

Similarly, from Eq.~(\ref{mtilde-elements}) one can see that 
the elements of $\mtilde$ are expected to be $\sim m_0^2$ 
in the absence of any cancellations. Hence if 
\be
{\rm Min}\left( \frac{|\mtilde(i,j)|}{m_0^2} \right) < 10^{-2}\,,
\label{zero:Mtilde}
\ee
then we conclude that the symmetry that requires $\mtilde(i,j)=0$ is 
still allowed. In the case of the discrete symmetries like $\mu-\tau$ 
exchange or $S_3$, certain ratios are expected to be equal to unity. 
Here, we demand that the deviation of such ratios from unity to be 
less than $10^{-2}$ for the symmetry to be acceptable.
Note that the right hand side of Eqs.~(\ref{zero:M}) and 
(\ref{zero:Mtilde}) can be changed to any small number of
one's choice, depending on how much deviation from the exact
symmetry one is willing to allow.
Our numerical results cover the complete relevant range, so
the required numbers can be read off from our figures.
 
Note that our criteria give the necessary conditions for a particular
symmetry to hold. Further considerations may disallow some of the
symmetry relations that are permitted by conditions in Eqs.~(\ref{zero:M})
and (\ref{zero:Mtilde}).

\section{Texture zeroes in $M_\nu$} 
\label{mnu-zeroes}

\subsection{Individual zeroes in $M_\nu$}
\label{mnu-onezero}

\begin{figure}[ht] 
\begin{center} 
\epsfig{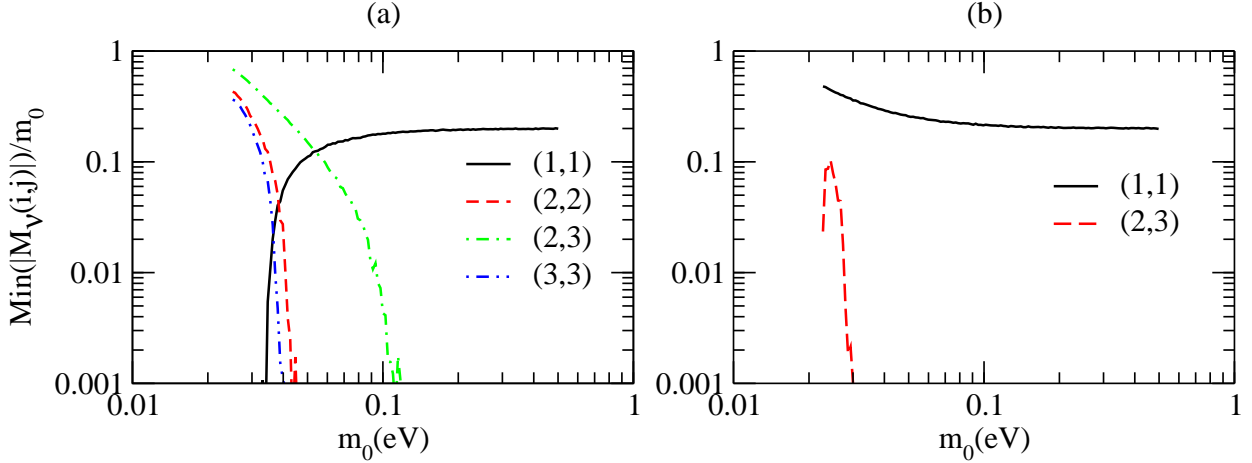} 
\caption{The minima of certain $M_\nu(i,j)/m_0$ are shown against $m_0$ for 
(a) normal and (b) inverted ordering of neutrino masses. The minima of the 
remaining elements of $M_\nu/m_0$ are less than 0.001 
for both normal as well as inverted ordering. 
The neutrino mixing parameters are varied in their $3\sigma$
allowed range in this and the subsequent figures.
\label{onezero-mnu}} 
\end{center} 
\end{figure} 

Whether a particular element in the neutrino mass matrix $M_\nu$
can potentially vanish can be checked analytically from Eqs.~(\ref{abcdef-def}).
To simplify the expressions, we define three quantities
\barr
\zeta_1 & \equiv & (1-\epsilon) e^{2i\phi_1} \; , \\
\zeta_2 & \equiv & (1+\epsilon) e^{2i\phi_2} \; ,\\
\zeta_3 & \equiv & s_{13} e^{i\delta} \; .
\label{zetas}
\earr
Note that since $\epsilon >0$, we have $|\zeta_1| < |\zeta_2|$ in case of NH, while  $|\zeta_1| 
\approx |\zeta_2| \approx 1$ in the IH and QD scenarios where $\epsilon \ll 1$.

The following observations may be made from the analytic expressions for 
$a, b, c, d, e, f$:

(i) \centerline{$a \equiv  M_\nu(1,1) = m_0 (1-\rho) e^{2 i \chi_e} 
\left[ \zeta_1 c_{12}^2 + \zeta_2 s_{12}^2 + (1+\rho) {\zeta_3^*}^2 \right] \;.$}

In NH, while the $(1-\rho)$ provides one suppression factor, the
other suppression is provided by the terms involving $\zeta_i$, all
of which have phases that can be adjusted so as to cause a 
cancellation among the three terms inside the square bracket.
Thus, $M_\nu(1,1)$ can vanish for NH as can be seen from Fig.~\ref{onezero-mnu}. 
Note that if $\theta_{13}$ were extremely small so as to make the $\zeta_3$
term negligible, the cancellation could not have been achieved since
current data implies $|\zeta_1 c_{12}^2| < |\zeta_2 s_{12}^2|$ in this
extreme hierarchical limit.
So a significantly non-vanishing $\theta_{13}$ is crucial for allowing 
a texture zero of $M_\nu(1,1)$ for NH.

In IH and QD, since $|\zeta_1|\approx |\zeta_2| \approx 1$ and 
$c_{12}^2 - s_{12}^2 > 0.26$, the minimum magnitude of the sum
of the first two terms in the square bracket is $\approx 0.26$. 
On the other hand, $|\zeta_3^2| < 0.04$, so this term 
cannot cancel the first two. Thus, the smallness of
$\theta_{13}$, combined with the non-maximality of $\theta_{12}$,
ensures that $M_\nu(1,1)$ cannot be a texture zero for IH and QD scenarios. 

In the normal mass ordering, extreme hierarchy allows the texture
zero of  $M_\nu(1,1)$ whereas quasidegeneracy prevents it.
The transition between these two extremes as a function of $m_0$
may be observed in Fig.~\ref{onezero-mnu}.

(ii) 
\centerline{$ b \equiv  M_\nu(1,2) \approx
- (m_0/\sqrt{2}) e^{i(\chi_e + \chi_\mu)} [\Omega_b  
+ {\cal O}(\theta_{13}^2,\tilde{\theta}_{23}) ] \; , $} 

$\phantom{(ii)}$ \centerline{$c \equiv  M_\nu(1,3) \approx
+ (m_0/\sqrt{2}) e^{i(\chi_e + \chi_\tau)} [\Omega_c  
+ {\cal O}(\theta_{13}^2,\tilde{\theta}_{23}) ] \; , $} 
where
$$\Omega_b = \Omega_c = 
(1-\rho) [c_{12} s_{12} (\zeta_1 - \zeta_2) - \zeta_3 (\zeta_1 c_{12}^2
+ \zeta_2 s_{12}^2)] + (1+\rho) \zeta_3^* \; .
$$ 
 
In NH, if $\theta_{13}=0$ then $\Omega_b$ and $\Omega_c$ 
cannot vanish, since $|\zeta_1| < |\zeta_2|$.
However, with a non-zero $\zeta_3$, the phases $\phi_1, \phi_2$ and $\delta$
may be chosen properly to make $\Omega_b$ and $\Omega_c$ vanish.
Note that the $(1+\rho)$ coefficient of the $\zeta_3^*$ term makes that
term comparable to the terms involving $\zeta_1,\zeta_2$, which are
suppressed by the coefficient $(1-\rho)$ for NH.
Thus, $M_\nu(1,2)$ and $M_\nu(1,3)$ can be texture zeroes in NH as long
as $\theta_{13}$ is not extremely small.

In IH and QD, $|\zeta_1| \approx |\zeta_2|$, so that 
$\phi_1 \approx \phi_2$ can make $\zeta_1-\zeta_2 \approx 0$. 
Thus even with $\theta_{13}$ vanishing, $M_\nu(1,2)$ and $M_\nu(1,3)$ can 
be texture zeroes of the neutrino mass matrix.

(iii) 
\centerline{$d \equiv  M_\nu(2,2) \approx
 (m_0/2) e^{2 i \chi_\mu} [\Omega_d + 
{\cal O}(\theta_{13}^2, \tilde{\theta}_{23})] \; , $}

$\phantom{(iii)}$ \centerline{ $f \equiv M_\nu(3,3) \approx 
 (m_0/2) e^{2 i \chi_\tau} [\Omega_f + 
{\cal O}(\theta_{13}^2, \tilde{\theta}_{23})] \; , $}
where
$$\Omega_d = \Omega_f = (1+\rho) + 
(1-\rho)[(\zeta_1 s_{12}^2 + \zeta_2 c_{12}^2) + 
2 \zeta_3 s_{12} c_{12} (\zeta_1-\zeta_2)] \; .
$$

In NH, $(1+\rho)$ is as high as 1.9. This cannot be cancelled by the 
other terms involving $\zeta_1, \zeta_2$ since these terms are
already suppressed by $(1-\rho)$.
As a result, $M_\nu(2,2)$ and $M_\nu(3,3)$ cannot be texture 
zeroes for NH. This can be seen from Fig.~\ref{onezero-mnu}(a). 
In IH, $(1+\rho) \approx 0$. While the coefficient of
$(1-\rho)$ in $\Omega_d$ and $\Omega_f$ 
cannot vanish for $\theta_{13}=0$, it can be made
to vanish for $\theta_{13} \neq 0$ with proper choices of
$\phi_1,\phi_2$ and $\delta$.
On the other hand in QD, the choice of $\phi_1 = \phi_2 = \pi/2$ would make
$\Omega_d$ and $\Omega_f$ vanish even when $\theta_{13}=0$.
Thus $M_\nu(2,2)$ and $M_\nu(3,3)$ can be texture zeroes in
IH and QD scenarios, the former scenario requiring
a nonzero $\theta_{13}$. 

\noindent(iv)
\centerline{ $e \equiv  M_\nu(2,3) \approx
 (m_0/2) e^{i (\chi_\mu+\chi_\tau)}  
[ (1+\rho)  - (1-\rho) (\zeta_1 s_{12}^2 + \zeta_2 c_{12}^2) +
{\cal O}(\theta_{13}^2, \tilde{\theta}_{23})] \; . $}

In NH, arguments similar to above can be used to show that the large
value of $(1+\rho)$ cannot be cancelled by the other terms that
are suppressed by $(1-\rho)$. As a result, $M_\nu(2,3)$ cannot
be a texture zero for NH.
In IH, while $(1-\rho)$ vanishes, the coefficient of $(1-\rho)$
term cannot vanish owing to the non-maximal nature of $\theta_{12}$,
preventing $M_\nu(2,3)$ from being a texture zero for IH. 
In QD, however,  $\phi_1 = \phi_2 = \pi/2$ would make $e$ vanish,
so that $M_\nu(2,3)$ can be a texture zero.

In Table~\ref{tab:mnuzero} we summarize our results on the viability 
of the texture zeroes in various scenarios.
\begin{table}
\begin{tabular}{lcccccc}
\hline
\phantom{space} & { $a$ } & { $b$ } & { $c$ } & { $d$ } & { $f$ } & { $e$ } \\
\hline
NH & $\Theta$ & $\Theta$  & $\Theta$ & $\otimes$ & $\otimes$ & $\otimes$ \\
IH & $\otimes$ & $\surd$ & $\surd$ & $\Theta$ & $\Theta$ & $\otimes$ \\
QD & $\otimes$ & $\surd$ & $\surd$ & $\surd$ & $\surd$ & $\surd$ \\
\hline
\end{tabular}
\caption{Viability of individual texture zeroes for $M_\nu$. 
Here $\surd$ indicates
that the texture is allowed even for $\theta_{13}$ vanishing,
$\Theta$ indicates that the texture is allowed but needs a 
nonzero $\theta_{13}$, whereas $\otimes$ indicates that the texture is
not allowed.
\label{tab:mnuzero}}
\end{table}

\subsection{Two zeroes in $M_\nu$} 
\label{mnu-twozero}

\begin{figure}
\begin{center}
\epsfig{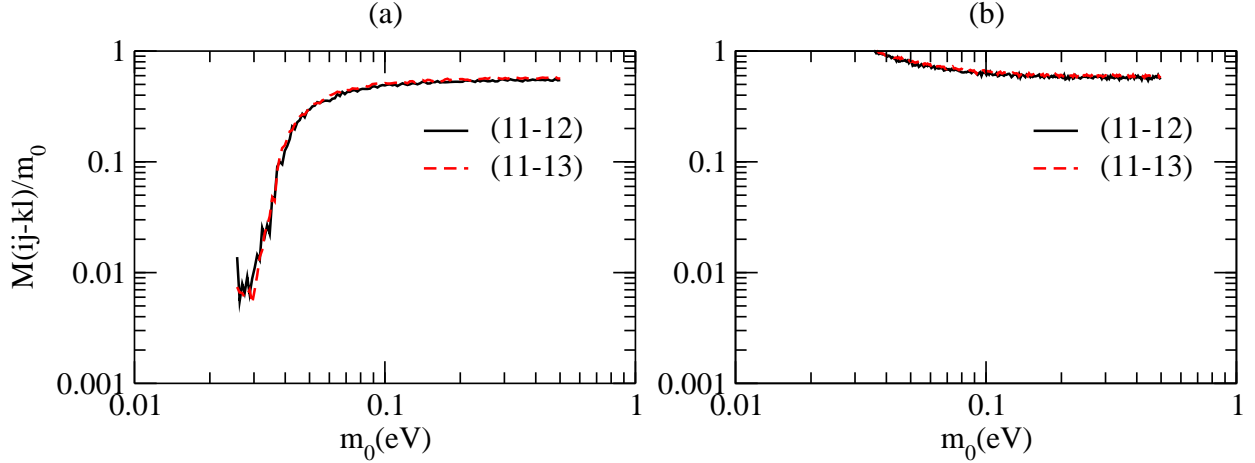}
\caption{$M(11-12)/m_0$ and $M(11-13)/m_0$ as
functions $m_0$ in (a) normal and (b) inverted ordering of neutrino 
masses. 
\label{old2zero_neu0}}
\end{center}
\end{figure}

In order to quantify the two-zero textures in $M_\nu$ in the bottom-up 
approach, we define 
\beq
M(ij-kl) \equiv {\rm Min} \left(
\frac{\sqrt{|M_\nu(i,j)|^2+|M_\nu(k,l)|^2}}{2} \right) \; .
\label{ijkl-def}
\eeq
There are fifteen $M(ij-kl)$ constructed out of six $M_\nu(i,j)$ elements.
From the discussion in Sec.~\ref{mnu-onezero}, 
it is clear that the five $M(ij-kl)$ involving 
$M_\nu(1,1)$ cannot vanish unless the mass ordering is normal and 
$m_0 \lsim 0.035 $ eV. Moreover, even in such a case the other element of 
the pair cannot be $M_\nu(2,2), M_\nu(2,3)$ or $M_\nu(3,3)$. The only 
two-zero textures including vanishing $M_\nu(1,1)$ are therefore
\barr
\begin{pmatrix} 0 & 0 & X\\  
0 & X & X\\ X & X & X\end{pmatrix} \quad & {\rm and} &  \quad 
\begin{pmatrix} 
0 & X & 0\\ X & X & X \\ 0 & X & X \end{pmatrix}\,,
\label{normal_textures} 
\earr
where $X$ denotes an element that may or may not vanish.
That both these patterns correspond to NH can
be seen from Fig.~\ref{old2zero_neu0}. 
The above two-zero textures with normal ordering imply that the neutrinoless 
double beta decay ($0\nu \beta \beta$), whose rate is proportional to 
\be
|M_\nu(1,1)|^2=\left| \sum_i m_i U_{1i}^2 \right|^2\,,
\ee
is unobservable since $|M_\nu(1,1)| < 10^{-2} m_0$~\cite{sanjeevetal.07}. 
In other words, if the neutrino masses are normally ordered and the 
$0\nu\beta \beta$ is  
observed then the corresponding symmetry which drives the cancellation 
in the neutrino mass matrix leading to these textures should be forbidden.

The five textures
\beq \label{twozero_notallowed}
\begin{pmatrix} X & 0 & 0\\ 0 & X & X \\ 0 & X & X \end{pmatrix}\,\, ,  
\begin{pmatrix} X & 0 & X\\ 0 & X & 0 \\ X & 0 & X\end{pmatrix}\,\, ,
\begin{pmatrix} X & X & 0 \\ X & X & 0 \\ 0 & 0 & X\end{pmatrix} \; ,
\begin{pmatrix} X & X & X \\ X & 0 & 0 \\ X & 0 & X\end{pmatrix}\,\,\,, 
\begin{pmatrix} X & X & X \\ X & X & 0\\ X & 0 & 0\end{pmatrix}\,\,\, 
\eeq
cannot satisfy the neutrino data since the first two lead to
$\theta_{12}=0$, the second and the third lead to $\theta_{23}=0$,
while the last two are inconsistent with a small $\theta_{13}$
and large $\theta_{23}$ simultaneously.

\begin{figure}
\begin{center}
\epsfig{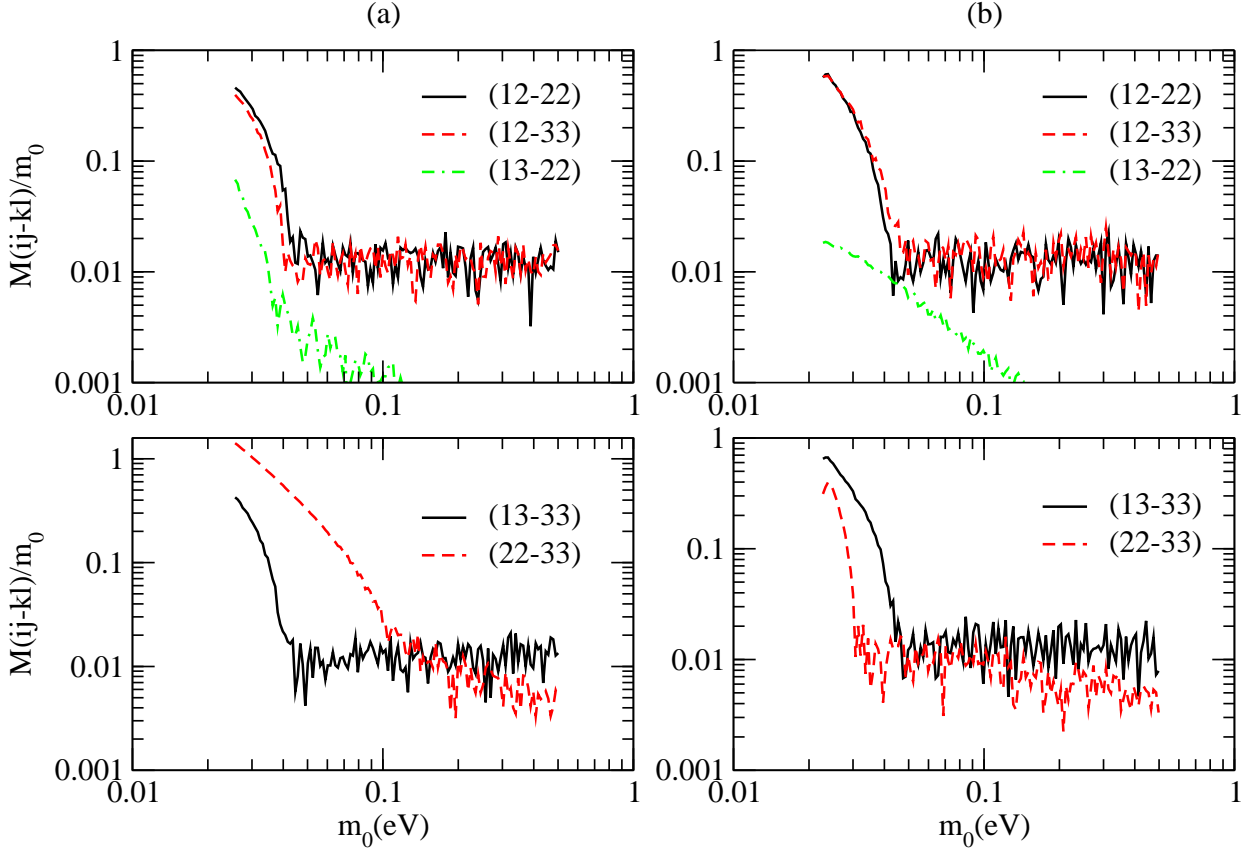}
\caption{$M(ij-kl)/m_0$ as functions of $m_0$ for (a) normal and 
(b) inverted ordering of neutrino masses. The fluctuations 
seen in this and some of the subsequent figures are
numerical artifacts, a consequence of the inability of the 
randomly chosen neutrino parameters to find the actual minimum.
\label{old2zeroneu1}}
\end{center}
\end{figure}

The remaining five pairs of $M_\nu(i,j)$ textures are allowed by both 
normal and inverted ordering of neutrino mass pattern at large $m_0$,
i.e. they are allowed in the QD scenario.
These textures are 
\begin{equation} 
\begin{pmatrix} X & 0 & X \\ 0 & 0 & X \\ X & X & X\end{pmatrix}\,\, , 
\begin{pmatrix} X & 0 & X \\ 0 & X & X\\ X & X & 0\end{pmatrix}\,\,, 
\begin{pmatrix} X & X & 0\\ X & 0 & X\\ 0 & X & X\end{pmatrix} \; ,
\begin{pmatrix} X & X & 0 \\ X & X & X\\ 0 & X & 0\end{pmatrix}\,\,, 
\begin{pmatrix} X & X & X\\ X & 0 & X \\ X & X & 0\end{pmatrix} \; .
\label{nor_inv_1}
\end{equation}
The $m_0$ values at which these textures start becoming viable
can be read off from Fig.~\ref{old2zeroneu1}.

Notice that the seven allowed two-zero textures, 
given by Eqs.~(\ref{normal_textures})  
and (\ref{nor_inv_1}), have already been predicted in the top-down scenario
\cite{top-down}. Here in addition to obtaining them through a 
bottom-up approach, we have also correlated them with the
measured values of $m_0$ and $\theta_{13}$.

\subsection{Three zeroes in $M_\nu$} 
\label{mnu-threezero}

\begin{figure}[htbp] 
\begin{center} 
\epsfig{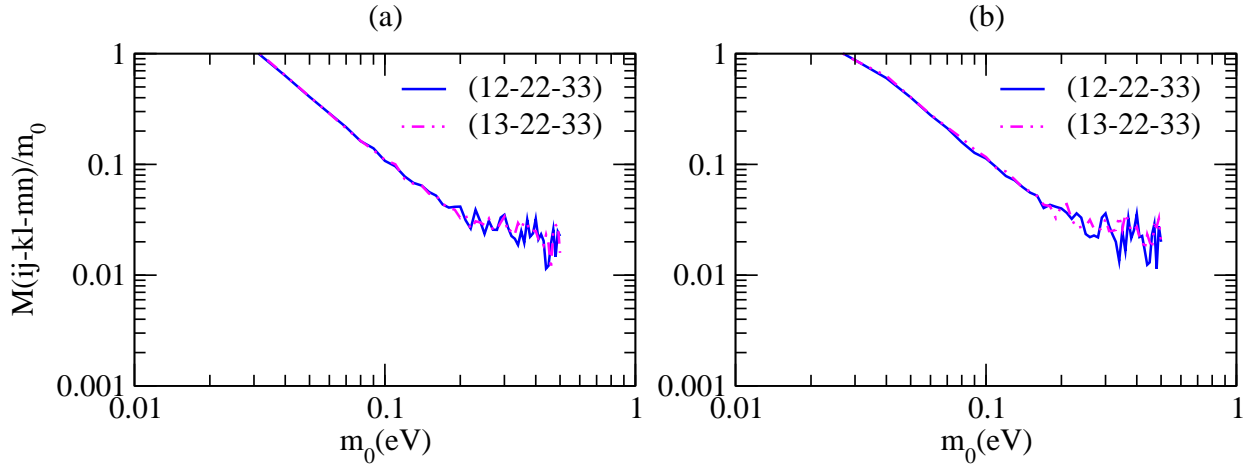}
\caption{$M(ij-kl-mn)/m_0$ as functions of $m_0$ for selected values 
of $(ij,kl,mn)$ for (a) normal and (b) inverted ordering of neutrino masses. 
For all the rest $(ij,kl,mn)$ values, 
$M(ij-kl-mn)$ is always greater than $0.1 m_0$ and hence 
the corresponding symmetry is forbidden.
\label{threezero-mnu}
}
\end{center} 
\end{figure}

In order to quantify the possible three-zero textures in $M_\nu$
in the bottom-up approach, we define 
\beq
M(ij-kl-mn) \equiv {\rm Min} \left(
\frac{\sqrt{|M_\nu(i,j)|^2+|M_\nu(k,l)|^2 + |M_\nu(m,n)|^2}}{3} \right) \; .
\label{ijklmn-def}
\eeq 
From the two-zero textures obtained in the last subsection, one can deduce 
that the only possible three-zero textures of $M_\nu$ are the combinations 
$(12-22-33)$ and $(13-22-33)$.
In Fig.~\ref{threezero-mnu}, we show these two
combinations $M(ij-kl-mn)/m_0$ as  functions of $m_0$ for normal as well as 
inverted ordering of neutrino masses.
It can be seen that $M(ij-kl-mn)/m_0$ does not vanish in any region
of the neutrino parameter space.
Thus, there are no viable three-zero textures for $M_\nu$. 

\section{Texture zeroes in $M_\nu^{-1}, \mtilde$ and $M_M$} 
\label{mM-zeroes}

The texture zeroes in $M_\nu^{-1}, \mtilde$ and $M_M$ are 
identical, as can be seen from Eqs.~(\ref{mnu-inv-def})
and (\ref{mM-def}). The quantification of ``smallness''
of the matrix elements is even more arbitrary for $M_M$
due to the presence of the unknown Dirac masses $x,y,z$.
However for the sake of uniformity, we determine the
texture zeroes of $\mtilde$ through the quantitative criteria
described in Sec.~\ref{whatiszero}, and apply the same
criteria for the zeroes of $M_M$. This correspondence
preserves all the zeroes in $\mtilde$, and does not add any
additional ones due to the hierarchy of Dirac masses, for
example.

We thus continue with determining the zeroes of $\mtilde$
on lines similar to the last section where we determined
zeroes of $M_\nu$. For convenience, we define
\barr
\mtilde(ij-kl)& \equiv & {\rm Min} \left(
\frac{\sqrt{|\mtilde(i,j)|^2+|\mtilde(k,l)|^2 }}{2} \right) \; ,
\label{tilde-ijkl-def} \\
\mtilde(ij-kl-mn) & \equiv & {\rm Min} \left(
\frac{\sqrt{|\mtilde(i,j)|^2+|\mtilde(k,l)|^2 + |\mtilde(m,n)|^2
}}{3} \right) \; .
\label{tilde-ijklmn-def}
\earr
The quantities ${\rm Min}\bigl(|\mtilde(i,j)| \bigr), 
\mtilde(ij-kl)$ and $\mtilde(ij-kl-mn)$
need to be less than $10^{-2} m_0^2$ in order to qualify as
one-, two- or three- zero textures respectively.

\subsection{Individual zeroes in $\mtilde$} 
\label{mM-onezero}

\begin{figure}[ht] 
\begin{center} 
\epsfig{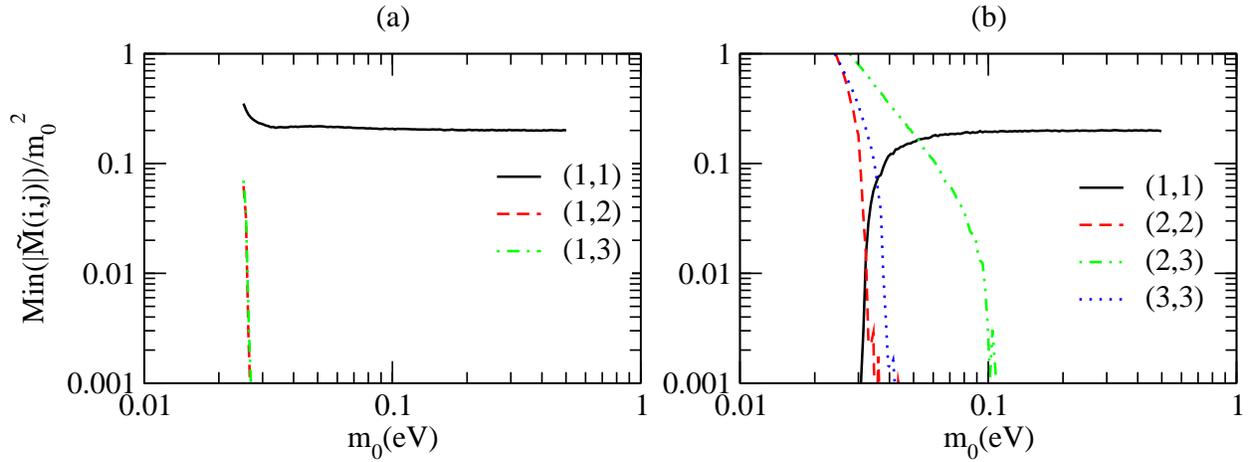}
\caption{Individual elements of Min$(\mtilde(i,j))/m_0^2$ as
functions of $m_0$ for (a) normal and (b) inverted ordering of neutrino 
mass spectrum. The minima of 
the remaining elements which are not shown in the figure are less than $0.001$ 
for all values of $m_0$.
\label{onezero-mtilde}} 
\end{center} 
\end{figure}

(i) \centerline{$A \equiv \mtilde(1,1) \approx m_0^2 e^{2i(\chi_\mu+\chi_\tau)} 
\left[ (1-\rho^2)  (\zeta_1 s_{12}^2 + \zeta_2 c_{12}^2) + 
(1-\rho)^2 \zeta_1 \zeta_2 \zeta_3^2 \right] \; .$}

In NH, the last term is suppressed quadratically in $(1-\rho)$ as
well as in $\zeta_3$, whereas the $(1-\rho^2)$ term does not have such a 
strong suppression. 
Moreover, the coefficient of $(1-\rho^2)$ cannot vanish since
$|\zeta_1 c_{12}^2| \ll |\zeta_2 s_{12}^2|$. 
As a result, $\mtilde(1,1)$ does not vanish for NH. 
In IH, $\rho^2 \approx 1$, so that the leading term vanishes. 
Further, the $(1-\rho)^2$ dependent term vanishes at $\theta_{13}=0$.
$\mtilde(1,1)$ as a texture zero is therefore allowed.
In QD on the other hand, the $(1-\rho^2)$ term and its coefficient are
nonzero, while the second term is suppressed by $(1-\rho)^2 \theta_{13}^2$
and is unable to cancel the first term.
As a result, $\mtilde(1,1)$ cannot vanish for QD.

In the inverted mass ordering, the hierarchical limit allows a texture
zero of  $\mtilde(1,1)$ whereas the quasidegenerate limit prevents it.
The transition between these two limits as a function of $m_0$
can be seen in Fig.~\ref{onezero-mtilde}(b).

(ii) \centerline{$ B \equiv \mtilde(1,2) \approx 
+ (m_0^2/\sqrt{2}) 
e^{i(\chi_e+\chi_\mu+2\chi_\tau)} [\Omega_B + 
{\cal O}(\theta_{13}^2, \tilde{\theta}_{23})] \; ,$ }

$\phantom{(ii)}$ \centerline{$ C \equiv \mtilde(1,3) \approx  
- (m_0^2/\sqrt{2}) 
e^{i(\chi_e+ 2\chi_\mu+ \chi_\tau)} [\Omega_C + 
{\cal O}(\theta_{13}^2, \tilde{\theta}_{23})] \; ,$ }
where
$$\Omega_B = \Omega_C =
(1-\rho^2)[ c_{12} s_{12} (\zeta_1 - \zeta_2) + 
\zeta_3 (\zeta_1 s_{12}^2 + \zeta_2 c_{12}^2 )]
-(1-\rho)^2 \zeta_1 \zeta_2 \zeta_3 \; . $$

In NH, the last term is suppressed by $(1-\rho)^2 \theta_{13}$, and is
therefore rather insignificant. 
However the coefficient of the $(1-\rho^2)$ term cannot vanish with
any choice of the Majorana and Dirac phases,
so $M_\nu(1,2)$ and $M_\nu(1,3)$ cannot be texture zeroes in NH. 
This feature can be seen from Fig.~\ref{onezero-mtilde}(a). 
In IH and QD, since $|\zeta_1| \approx |\zeta_2|$, one may choose 
$\phi_1 \approx \phi_2$ to make $\zeta_1-\zeta_2 \approx 0$,
so that $\mtilde(1,2)$ and $\mtilde(1,3)$ can be texture zeroes.

(iii) 
\centerline{$ D \equiv \mtilde(2,2) \approx m_0^2 e^{2i(\chi_e+\chi_\tau)} 
[\Omega_D + {\cal O}(\theta_{13}^2, \tilde{\theta}_{23}^2) ] \; ,$}

$\phantom{(iii)}$  \centerline{$F \equiv \mtilde(2,3) \approx 
m_0^2 e^{2i(\chi_e+\chi_\mu)} 
[\Omega_F + {\cal O}(\theta_{13}^2, \tilde{\theta}_{23}^2) ] \; ,$}

$\phantom{(iii)}$  \centerline{$E \equiv \mtilde(2,3) 
\approx m_0^2 e^{i(2\chi_e+\chi_\mu+\chi_\tau)} 
[\Omega_E + {\cal O}(\theta_{13}^2, \tilde{\theta}_{23}^2) ] \; ,$}

where
\barr
\Omega_D & = & \frac{(1-\rho)^2}{2} \zeta_1 \zeta_2 +
\frac{(1-\rho^2)}{2} (\zeta_1 c_{12}^2+\zeta_2 s_{12}^2 - 2
\zeta_3^* (\zeta_1-\zeta_2) c_{12} s_{12} ) \; , \nonumber \\
\Omega_F & = & \frac{(1-\rho)^2}{2} \zeta_1 \zeta_2 +
\frac{(1-\rho^2)}{2} (\zeta_1 c_{12}^2+\zeta_2 s_{12}^2 + 2
\zeta_3^* (\zeta_1-\zeta_2) c_{12} s_{12} )  \; , \nonumber \\
& & - 2 (1-\rho)^2 c_{12} s_{12} \zeta_1 \zeta_2 \zeta_3 
\nonumber \\
\Omega_E & = & - \frac{(1-\rho)^2}{2} \zeta_1 \zeta_2 +
\frac{(1-\rho^2)}{2} (\zeta_1 c_{12}^2+\zeta_2 s_{12}^2 + 2
\zeta_3^* (\zeta_1-\zeta_2) c_{12} s_{12} ) \; . \nonumber 
\earr

In NH, the first term in each of the expressions $\Omega_D, \Omega_F, \Omega_E$
is suppressed by $(1-\rho)^2 \zeta_1$,
whereas the coefficient of $(1-\rho^2)$ can be made to vanish
with an appropriate choice  of phases, as long as $\theta_{13}$
is finite and $\zeta_3$ can participate in the cancellation.
So $\mtilde(2,2),\mtilde(2,3)$ and $\mtilde(3,3)$ can be texture zeroes for NH.
In IH, though the $(1-\rho^2)$ term vanishes, the
coefficient of the $(1-\rho)^2$ term is of the order of unity
and as a result,
none of the above three can be texture zeroes.
In QD, the first term is suppressed by $(1-\rho)^2 \zeta_1$,
while the coefficient of $(1-\rho^2)$ becomes
$$ (\zeta_1 c_{12}^2+\zeta_2 s_{12}^2 \pm \zeta_1 \zeta_2)\;$$
when $\theta_{13}=0$.
The choices of the Majorana phases $\phi_1 = \phi_2 = \pi/2$ 
and $\phi_1 = \phi_2 = 0$ make the above expression vanish for
the + and - sign respectively.
Thus, $M_\nu(2,2), M_\nu(2,3)$ and $M_\nu(3,3)$
are allowed as texture zeroes for QD.

In the inverted mass ordering, the hierarchical limit allows a texture
zero of the above three matrix elements, whereas the quasidegenerate 
limit prevents it.
The transition between these two limits can be seen in
Fig.~\ref{onezero-mtilde}(b).

In Table~\ref{tab:mMzero} we summarize our results on the viability 
of the texture zeroes in various scenarios.

\begin{table}
\begin{tabular}{lcccccc}
\hline
\phantom{space} & { $A$ } & { $B$ } & { $C$ } & { $D$ } & { $F$ } & { $E$ } \\
\hline
NH & $\otimes$ & $\otimes$  & $\otimes$ & $\Theta$ & $\Theta$ & $\Theta$ \\
IH & $\surd$ & $\surd$ & $\surd$ & $\otimes$ & $\otimes$ & $\otimes$ \\
QD & $\otimes$ & $\surd$ & $\surd$ & $\surd$ & $\surd$ & $\surd$ \\
\hline
\end{tabular}
\caption{Viability of individual texture zeroes for $M_M$. Here
$\surd$ indicates
that the texture is allowed even for $\theta_{13}$ vanishing,
$\Theta$ indicates that the texture is allowed but needs a 
nonzero $\theta_{13}$, whereas $\otimes$ indicates that the texture is
not allowed.
\label{tab:mMzero}}
\end{table}

\subsection{Two zeroes in $\mtilde$} 
\label{mM-twozero}

Out of fifteen elements in $\widetilde{M}(ij-kl)$, 
five two-zero textures involve a zero of $\mtilde(1,1)$. 
These are clearly not allowed for NH and QD, since the value 
of $\mtilde(1,1)$ itself is high in these scenarios.
In IH, $\mtilde(1,1)$ can be small, however 
$\mtilde(2,2), \mtilde(2,3)$ and $\mtilde(3,3)$ are large,
so that $\mtilde(11-22), \mtilde(11-23)$ and $\mtilde(11-33)$
cannot be texture zeroes. This leaves us with the two
possible textures
\begin{equation} 
\begin{pmatrix} 0 & 0 & X\\ 0 & X & X\\ X & X & X\end{pmatrix}\,\, , 
\begin{pmatrix} 0 & X & 0\\ X & X & X\\ 0 & X & X\end{pmatrix}\,\, .
\label{two_left}\\ 
\end{equation}
However, these two inverse neutrino mass matrices correspond to the last two
two-zero textures of $M_\nu$ matrices in (\ref{twozero_notallowed}), 
which are forbidden. 
Thus, even the textures in (\ref{two_left}) are not allowed.

The textures
\beq
\begin{pmatrix} X & 0 & 0\\ 0 & X & X \\ 0 & X & X\end{pmatrix}\,\, , 
\begin{pmatrix} X & 0 & X\\ 0 & X & 0 \\ X & 0 & X\end{pmatrix}\,\, , 
\begin{pmatrix} X & X & 0\\ X & X & 0 \\ 0 & 0 & X\end{pmatrix} \; 
\label{twozero_ineu_3} 
\eeq
are ruled out since in each of these, one of the neutrinos is completely
decoupled, whereas we need large values for 
$\theta_{12}$ as well as $\theta_{23}$.

\begin{figure}
\begin{center} 
\epsfig{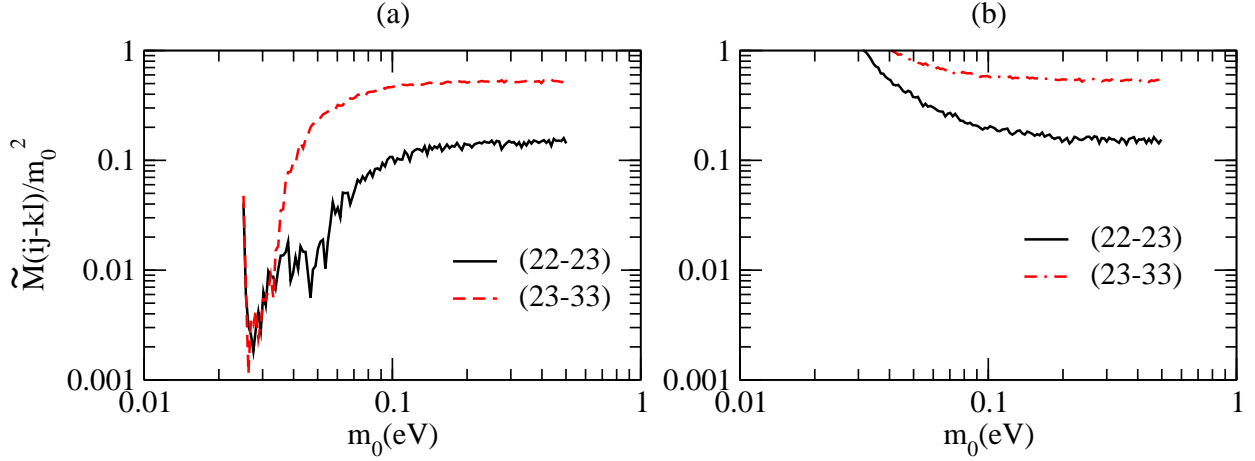}
\caption{Minima of $\mtilde(22-23)$ and $\mtilde(23-33)$ as
functions of  $m_0$ for 
(a) normal and (b) inverted ordering of neutrino masses.
\label{twozero-mtilde-1}
}
\end{center} 
\end{figure}

Textures involving zeroes of $\mtilde(2,2), \mtilde(2,3)$ 
and $\mtilde(3,3)$ can be allowed only for NH and QD.
It is observed that $\mtilde(22-23)$ and $\mtilde(23-33)$ 
can vanish in the NH scenario, but not in the rest.
The dependence of the vanishing of these two textures,
\begin{eqnarray}
\begin{pmatrix} X & X & X \\ X & X & 0\\ X & 0 & 0\end{pmatrix} \; ,
\begin{pmatrix} X & X & X \\ X & 0 & 0 \\ X & 0 & X\end{pmatrix}\, ,
\label{normal_adj_texture}
\end{eqnarray}
can be seen from Fig.~\ref{twozero-mtilde-1} where we have plotted 
$\mtilde(ij-kl)/m_0^2$ as a function of $m_0$ for normal as well as inverted 
ordering of neutrino masses. 

\begin{figure}
\begin{center}
\epsfig{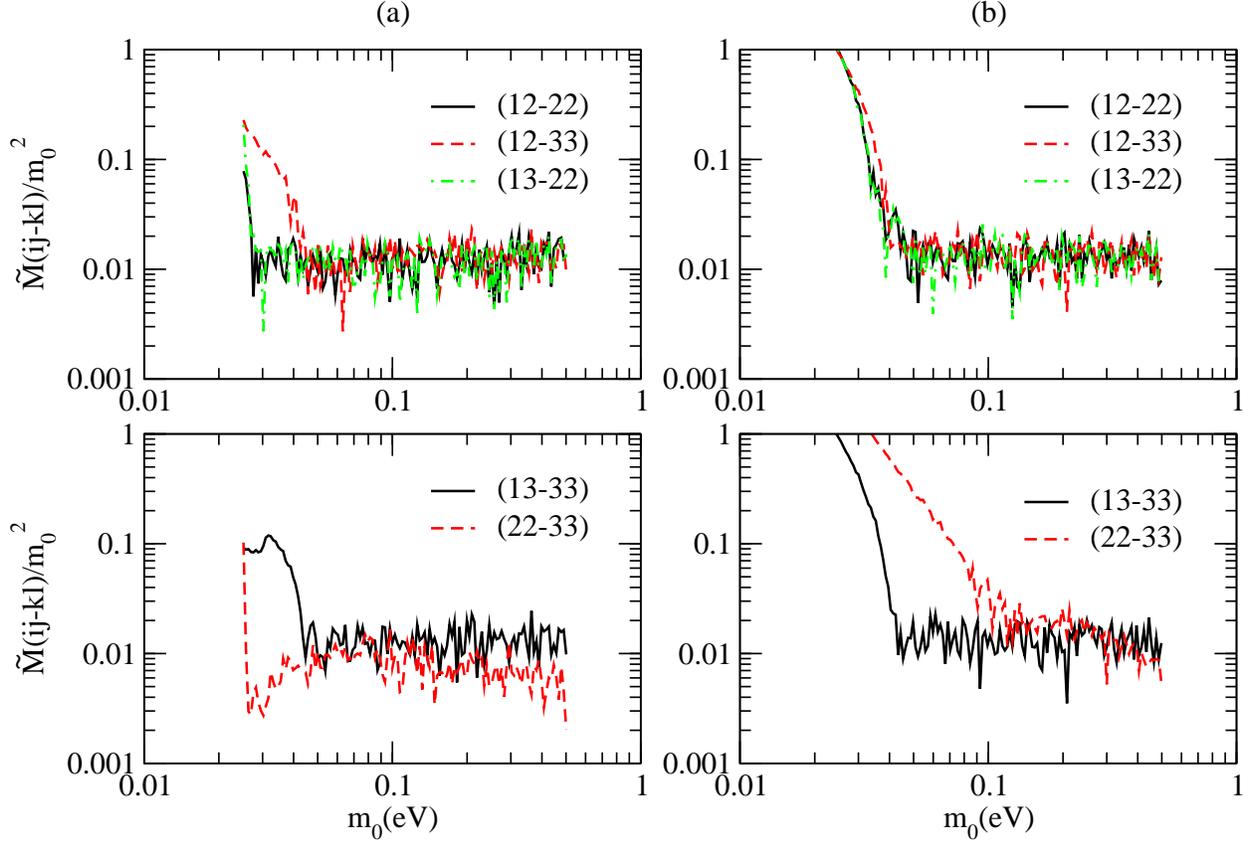}
\caption{Selected $\mtilde(ij-kl)/m_0^2$ as functions of $m_0$ 
for (a) normal  and (b) inverted ordering of neutrino masses.
\label{twozero-mtilde-2}}
\end{center}
\end{figure}

The remaining five two-zero textures, 
\begin{equation} 
\begin{pmatrix} X & 0 & X\\ 0 & 0 & X\\ X & X & X\end{pmatrix}\,\, , 
\begin{pmatrix} X & 0 & X\\ 0 & X & X\\ X & X & 0\end{pmatrix}\,\, , 
\begin{pmatrix} X & X & 0\\ X & 0 & X\\ 0 & X & X\end{pmatrix} \; ,
\begin{pmatrix} X & X & 0\\ X & X & X\\ 0 & X & 0\end{pmatrix}\; 
{\rm and}
\begin{pmatrix} X & X & X \\ X & 0 & X\\ X & X & 0\end{pmatrix} \; ,
\label{twozero_ineu_1}\\ 
\end{equation}
are allowed by both normal and  
inverted ordering of neutrino mass pattern, 
though only in the quasi-degenerate limit. This
can be seen quantitatively in Fig.~\ref{twozero-mtilde-2}.
Note that the textures given in Eqs. (\ref{normal_adj_texture}),  
and (\ref{twozero_ineu_1}) are already predicted in the 
top-down scenario \cite{top-down}. 
Here in addition to obtaining them through a 
bottom-up approach, we have also correlated them with the
measured values of $m_0$ and $\theta_{13}$.

\subsection{Three zeroes in $\mtilde$} 
\label{mM-threezero}

%
\begin{figure}[htbp] 
\begin{center} 
\epsfig{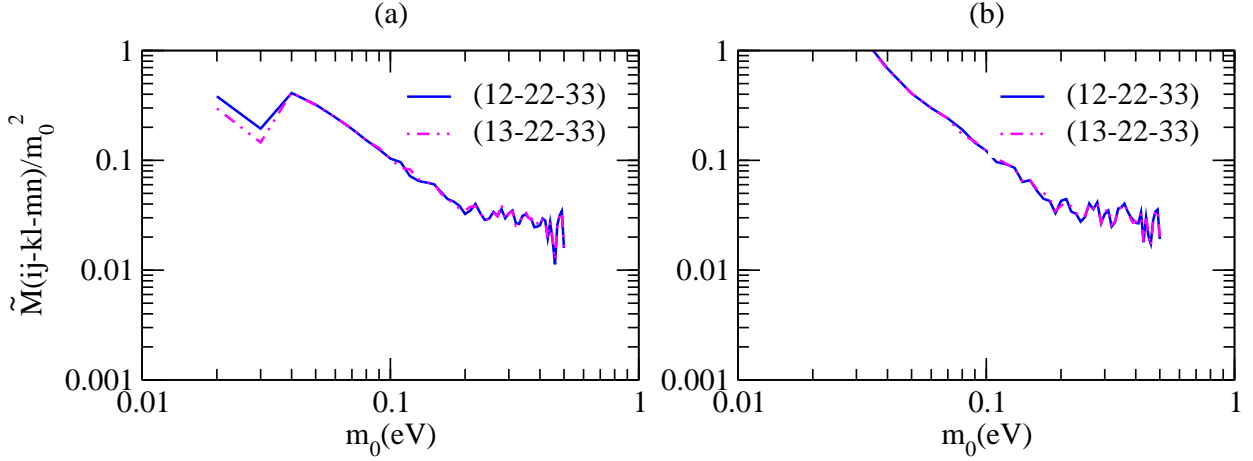}
\caption{$\mtilde(ij-kl-mn)/m_0^2$ as  functions of $m_0$, for
potentially viable values of $(ij,kl,mn)$ in case of (a) normal as well as 
(b) inverted ordering of neutrino masses.
\label{threezero-mtilde}
}
\end{center} 
\end{figure}
Using the two-zero textures of $\widetilde{M}$ predicted in the last
section, there are only two possible three-zero textures that are
viable: $(12-22-33)$ and $(13-22-33)$. Fig.~\ref{threezero-mtilde} 
shows the value of $\mtilde(ij-kl-mn)/m_0^2$ for these two
cases.
As can be seen from the figure, there are no three-zero
textures of the adjoint neutrino mass matrix $\mtilde$, and hence
of $M_M$.

\section{$\mu-\tau$ and $S_3$ symmetries in $M_\nu$}
\label{mu-tau}

\subsection{$\mu - \tau$ exchange symmetry in $M_\nu$} 
\label{mnu-mutau}

A $\mu-\tau$ symmetry in $M_\nu$ implies that, in the notation of 
(\ref{mnu-def}), 
we have $d=f$ and $b=c$. The deviations from $\mu-\tau$ symmetry may be then 
quantified in terms of the dimensionless parameter   
\begin{equation} 
\Delta_{\mu\tau}=\left|1-\frac{d}{f}\right|^2 + \left|1-\frac{b}{c}\right|^2\,. 
\end{equation} 
The existence of a $\mu-\tau$ symmetry in $M_\nu$, as per our convention
in Sec.~\ref{whatiszero}, requires 
Min$(\Delta_{\mu\tau}) < 10^{-2}$.
In order to analytically understand the constraints on various parameters,
we expand $\Delta_{\mu\tau}$ in terms of the relevant set of small
parameters.

For NH, using Eq.~(\ref{abcdef-def}) one can calculate
\barr
(1 - d/f)_{\rm NH} & = &
1 - e^{2 i(\chi_\mu - \chi_\tau)} 
+ {\cal O}(\tilde\rho,\tilde\epsilon,\theta_{13},\tilde\theta_{23}) \; ,
\nonumber \\
(1- b/c)_{\rm NH} & = &
1 - e^{i(\chi_\mu -\chi_\tau)} \left(
\frac{2 \theta_{13} + e^{i(2 \phi_2 +\delta)}
\tilde\rho \sin(2 \theta_{12})}{2 \theta_{13} - e^{i(2 \phi_2 +\delta)}
\tilde\rho \sin(2 \theta_{12})}
\right)
+ {\cal O}(\tilde\rho,\tilde\epsilon,\theta_{13},\tilde\theta_{23})
\; .
\label{bc-nh}
\earr
For $\Delta_{\mu\tau} \approx 0$, we need both the above expressions to
vanish, which can only happen if
$|\chi_\mu - \chi_\tau| \approx \pi$ and 
$\theta_{13} \ll \tilde\rho \sin(2 \theta_{12})$.

In the IH scenario, we get
\barr
(1 - d/f)_{\rm IH} & = &
1 - e^{2 i(\chi_\mu - \chi_\tau)} 
+ {\cal O}(\widehat\rho, \epsilon,\theta_{13},\tilde\theta_{23}) \; ,
\nonumber \\
(1- b/c)_{\rm IH} & = &
1 + e^{i(\chi_\mu -\chi_\tau)} 
+ {\cal O}(\widehat\rho,\epsilon,\theta_{13},\tilde\theta_{23}) \;
\; ,
\label{bc-ih}
\earr
so that $|\chi_\mu - \chi_\tau| \approx \pi$ is enough to get
$\Delta_{\mu\tau} \approx 0$ to zeroth order in the relevant
small parameter.

\begin{figure}[ht]
\begin{center}
\epsfig{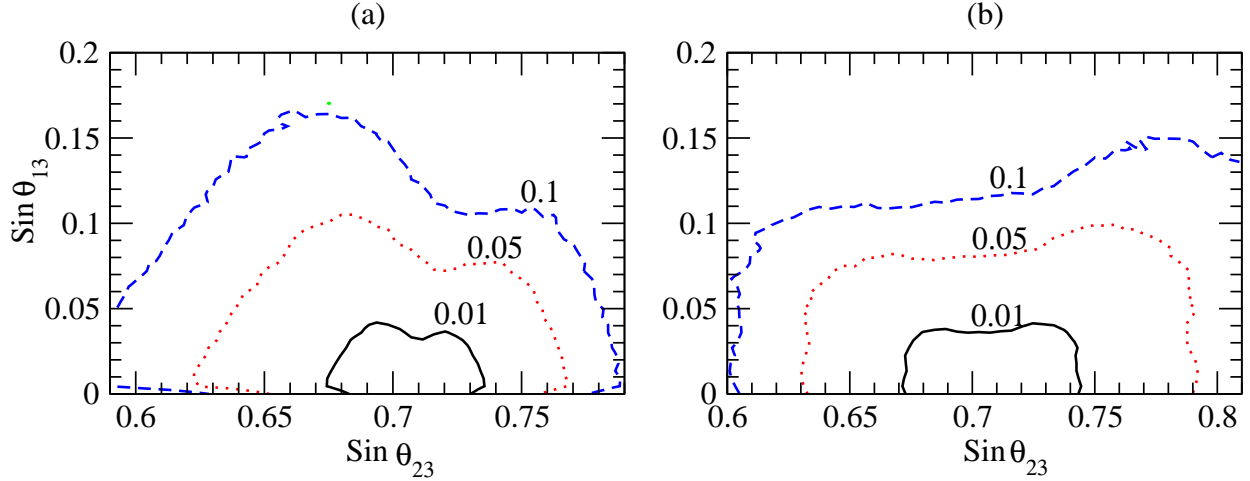}
\caption{The contours of $\Delta_{\mu\tau}$, the deviation from the $\mu-\tau$ 
exchange symmetry in $M_\nu$, in the plane of $\sin \theta_{23}$ versus 
$\sin\theta_{13}$ for (a) normal and (b) inverted mass ordering,
for $m_0 = 0.03$ eV.
\label{mutausymmetry_neu}}
\end{center}
\end{figure}

The QD scenario also gives 
\barr
(1 - d/f)_{\rm QD} & = &
1 - e^{2 i(\chi_\mu - \chi_\tau)} 
+ {\cal O}(\tilde\rho,\tilde\epsilon,\theta_{13},\tilde\theta_{23}) \; , 
\nonumber \\
(1- b/c)_{\rm QD} & = &
1 + e^{i(\chi_\mu -\chi_\tau)} 
+ {\cal O}(\tilde\rho,\tilde\epsilon,\theta_{13},\tilde\theta_{23}) \; , 
\label{bc-qd}
\earr
so that we expect $\Delta_{\mu\tau} \approx 0$ to be satisfied
if $|\chi_\mu - \chi_\tau| \approx \pi$.

In Fig.~\ref{mutausymmetry_neu}, we show the numerical results for 
Min$(\Delta_{\mu\tau})$ for hierarchical neutrino masses with
both mass orderings.
The current data allow for exact $\mu-\tau$ symmetry 
whatever the value of $m_0$ or nature of mass ordering,
 it will be possible to rule
out this symmetry if a significant value of $\theta_{13}$ 
is measured at experiments.

\subsection{$S_3$ permutation symmetry in $M_\nu$}
\label{mnu-s3}

The $S_3$ permutation symmetry in $M_\nu$ demands that, in addition to 
the conditions $d=f$ and $b=c$ satisfied by the $\mu-\tau$ exchange symmetry, 
one also needs $b=e$ and $a=f$. 
The deviation from $S_3$ symmetry then can be quantified by the 
dimensionless parameter
\begin{equation}
\Delta_{S_3} \equiv \left|1-\frac{d}{f}\right|^2+\left|1-\frac{b}{c}\right|^2
+\left|1-\frac{a}{f}\right|^2+ \left|1-\frac{b}{e}\right|^2\,.
\end{equation}

Clearly, the $\mu-\tau$ symmetry is a subset of the $S_3$ permutation
symmetry. Then the discussion in Sec.~\ref{mnu-mutau} shows that 
we need the conditions in Eqs.~(\ref{bc-nh}), (\ref{bc-ih}) and
(\ref{bc-qd}) satisfied in the NH, IH and QD scenarios
respectively, to ensure $|1-d/f| \ll 1$ and $|1-b/c| \ll 1$. 
In addition, we need to check if
the terms $|1-a/f|$ and $|1-b/e|$ can be small quantities.
We shall again analyze the relevant parameter space by considering
the analytic expansion in the relevant small parameters as done
in the previous section.

In the NH scenario,
\barr
(1 - a/f)_{\rm NH} & = & 1 
+ {\cal O}(\tilde\rho,\tilde\epsilon,\theta_{13},\tilde\theta_{23})
\; , \nonumber \\
(1 - b/e)_{\rm NH} & = & 1 
+ {\cal O}(\tilde\rho^2,\tilde\epsilon^2,\theta_{13}^2,\tilde\theta_{23}^2)
\; . 
\earr
None of the above conditions can be satisfied in this region of parameter
space, where $\tilde\rho \ll 1$. Therefore, one cannot have $S_3$
symmetry.

In the IH scenario,
\barr
(1 - a/f)_{\rm IH} & = & 1 - 2 e^{2i(\chi_e-\chi_\tau)}
\left( \frac{e^{2 i \phi_1} c_{12}^2 + e^{2 i \phi_2} s_{12}^2}
{e^{2 i \phi_1} s_{12}^2 + e^{2 i \phi_2} c_{12}^2} \right)  
+ {\cal O}(\widehat\rho,\epsilon,\theta_{13},\tilde\theta_{23}) \; . \\
(1 - b/e)_{\rm IH} & = & 1 - \sqrt{2} e^{i(\chi_e-\chi_\tau)}
\left( \frac{(e^{2 i \phi_1} - e^{2 i \phi_2}) c_{12} s_{12}}
{e^{2 i \phi_1} s_{12}^2 + e^{2 i \phi_2} c_{12}^2}  \right)
+ {\cal O}(\widehat\rho,\epsilon,\theta_{13},\tilde\theta_{23}) \; .
\earr
These conditions correspond to strict correlations between
the flavor phases, the Majorana phases and the mixing angle
$\theta_{12}$.
However, the above two equations cannot be satisfied simultaneously.
This can be seen as follows: the simultaneous conditions 
$a/f \approx 1$ and $b/e\approx 1$ would imply $(b/e)^2 \approx a/f$,
or $(b/e)^2 - a/f \approx 0$. However, one gets
\beq
[(b/e)^2 - a/f]_{\rm IH} = \frac{2 e^{2 i (\chi_e - \chi_\tau + \phi_1 - \phi_2)}}{
(e^{2 i \phi_2} c_{12}^2 + e^{2 i \phi_1} s_{12}^2)^2 }
+ {\cal O}(\widehat\rho,\epsilon,\theta_{13},\tilde\theta_{23}) \; ,
\eeq
which does not vanish even to the zeroth order of the small
parameters.
Thus, the current data do not allow $S_3$ symmetry to hold
in the IH scenario.

On the other hand, when neutrinos are quasi-degenerate in mass,
\barr
(1 - a/f)_{QD} & = & 1 - 2 e^{2i(\chi_e-\chi_\tau)}
\left( \frac{e^{2 i \phi_1} c_{12}^2 + e^{2 i \phi_2} s_{12}^2}
{1 + e^{2 i \phi_1} s_{12}^2 + e^{2 i \phi_2} s_{12}^2}  \right)
+ {\cal O}(\tilde\rho,\tilde\epsilon,\theta_{13},\tilde\theta_{23}) \; . \\
(1 - b/e)_{QD} & = & 1 + \sqrt{2} e^{i(\chi_e-\chi_\tau)}
\left( \frac{(e^{2 i \phi_1} - e^{2 i \phi_2}) c_{12} s_{12}}
{1 - e^{2 i \phi_1} s_{12}^2 - e^{2 i \phi_2} s_{12}^2} \right)  
+ {\cal O}(\tilde\rho,\tilde\epsilon,\theta_{13},\tilde\theta_{23}) \; .
\earr
which also constrain $\theta_{12}, \phi_1,\phi_2,\chi_e, \chi_\tau$.
The above two equations have simultaneous solutions, so the current 
data allow $S_3$ symmetry to hold in the QD scenario.

\begin{figure}[ht]
\begin{center}
\epsfig{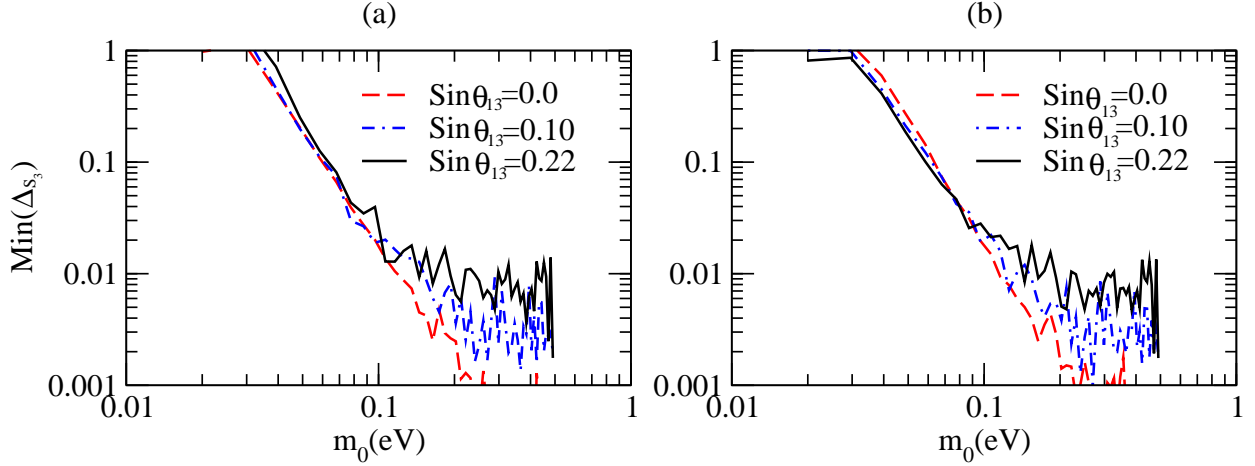}
\caption{The deviation of $S_3$ symmetry in $M_\nu$ as a function of $m_0$ for
(a) normal and (b) inverted mass ordering.
\label{strongs3_neu}}
\end{center}
\end{figure}

\begin{figure}[ht]
\begin{center}
\epsfig{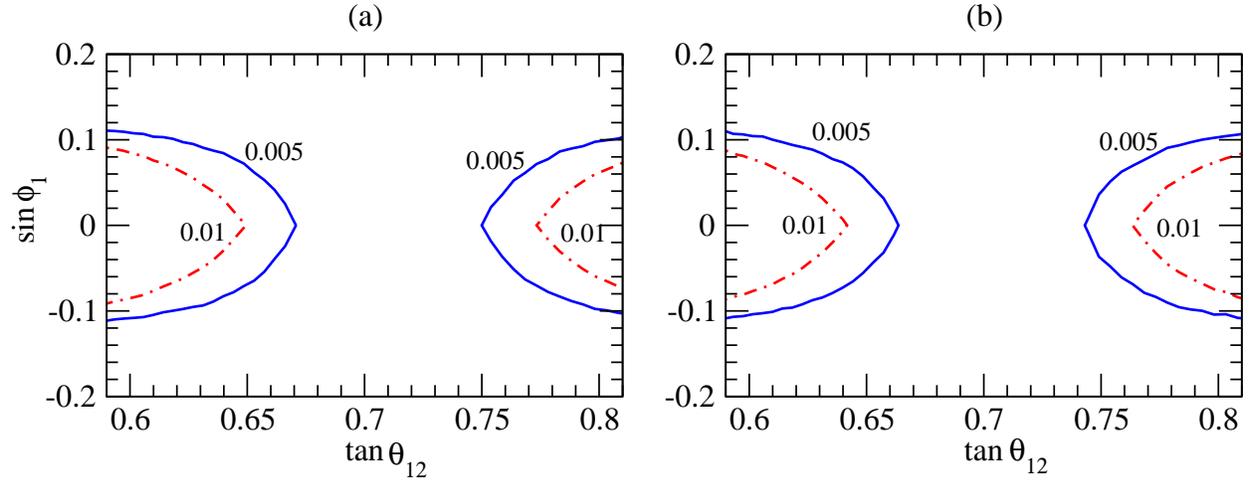}
\caption{The contours of ${\Delta}_{S_3}$, deviation of $S_3$ 
permutation symmetry in $M_\nu$, is shown in the 
$\tan\theta_{12}$--$\sin\phi_1$ plane.
for (a) normal and (b) inverted  
mass ordering, and $m_0 = 0.2$ eV.
\label{th12-phi1}}
\end{center}
\end{figure}

Thus, the $m_0$ value determines whether $S_3$ is an allowed 
symmetry in the normal as well as inverted ordering.
Fig.~\ref{strongs3_neu} shows the transition between hierarchical and 
degenerate scenarios where $S_3$ becomes viable.
In Fig.~\ref{th12-phi1}, we show the deviation from $S_3$ symmetry,
which we have quantified in terms of $\Delta_{S_3}$, as a function of
$\theta_{12}$ and $\phi_1$. Since these two parameters can be directly
constrained from the neutrinoless double beta decay, these
experiments will be crucial for testing the $S_3$ symmetry
in neutrinos.

\section{$\mu - \tau$ and $S_3$ symmetries in $M_M$}
\label{mM} 

\subsection{$\mu-\tau$ symmetry in $M_M$}

A $\mu -\tau$ exchange symmetry in $M_M$ implies, from (\ref{mM-def}),
that $xyB = xzC$ and $y^2 D = z^2 F$. Since we are completely
ignorant about the Dirac masses $x,y,z$, we cannot use these two
conditions separately. However, we can use the Dirac-mass independent
combination of these conditions, which gives  $(B/C)^2=(D/F)$. 
The deviation from this $\mu-\tau$ symmetry relation may be quantified by
\begin{equation} 
\widetilde\Delta_{\mu\tau} \equiv \left|\frac{B^2}{C^2}-\frac{D}{F}\right|^2 \,. 
\label{v-eqn}
\end{equation} 

Since this condition is less restrictive than that for the $\mu$-$\tau$
symmetry in $M_\nu$, which itself is consistent with the current data
irrespective of the value of $m_0$, one would expect that the
$\mu$-$\tau$ symmetry will hold also for $\widetilde{M}$.
Indeed, it is found that Min$(\widetilde\Delta_{\mu\tau})\ll 10^{-2}$ 
for all allowed values of the neutrino parameters. 
This can be understood as follows. In the NH scenario,
the expansion in terms of the small parameters $\theta_{13}, 
\tilde\theta_{23}, \tilde\epsilon, \tilde\rho$ gives
\beq
(B^2/C^2-D/F)_{\rm NH} = 
 {\cal O}(\tilde\rho^2,\tilde\epsilon^2,\theta_{13}^2,\tilde\theta_{23}^2) \; ,
\label{mutau-NH}
\eeq
where we have used Eq.~(\ref{mtilde-elements}).
Note that even terms linear in the small parameters $\theta_{13}, 
\tilde\theta_{23}, \tilde\epsilon, \tilde\rho$ are absent. 
This indicates that in this scenario, the $\mu-\tau$ exchange symmetry
is easily satisfied.
In the IH scenario,
\beq
(B^2/C^2 - D/F)_{\rm IH} =-8e^{-2i(\chi_\mu -\chi_\tau)} \tilde{\theta}_{23}+
 \frac{8 \theta_{13} e^{i(2\chi_\mu - 2 \chi_\tau - 2\phi_1 - 2\phi_2 -\delta)} }{c_{12} s_{12} 
\widehat{\rho} (e^{2i\phi_1} - e^{2 i \phi_2})} \;,
\label{mutau-IH}
\eeq
which vanishes for $\tilde{\theta}_{23}, \theta_{13}\rightarrow 0$. 
In the QD scenario also, $\mu-\tau$ exchange symmetry is easily
satisfied since
\beq
(B^2/C^2-D/F)_{\rm QD} = 
 {\cal O}(\tilde\rho,\epsilon,\theta_{13},\tilde\theta_{23}) \; .
\label{mutau-QD}
\eeq
Thus the $\mu-\tau$ symmetry is consistent with the current data
irrespective of $m_0$ or the mass ordering of neutrinos.

\begin{figure}[htbp]
\begin{center}
\epsfig{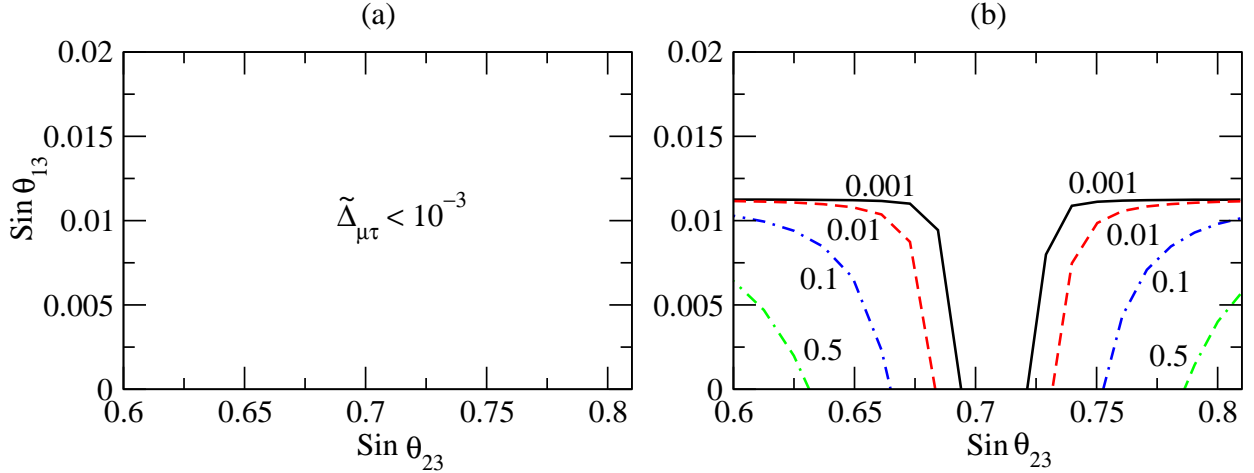}
\caption{The contours of $\tilde{\Delta}_{\mu\tau}$, 
the deviation from the $\mu$--$\tau$ permutation symmetry in $M_M$, in 
the plane of $\sin \theta_{23}$ versus $\sin \theta_{13}$ for (a) normal and (b) inverted mass ordering, for 
$m_0=0.025$ eV. 
\label{IH_theta23}}
\end{center}
\end{figure}

In the IH scenario, the deviation from the $\mu$--$\tau$ symmetry 
would manifest itself in the angles $\theta_{13}$ and $\theta_{23}$.
Clearly the symmetry is valid if $\theta_{13}=0=\tilde\theta_{23}$.
If $\theta_{13}=0$ exactly and $\tilde\theta_{23} \neq 0$, 
Eq.~(\ref{mutau-IH}) indicates that $(B^2/C^2 - D/F)_{\rm IH}$ 
cannot vanish, thus the symmetry is not obeyed.
However, if the upper bound on $\theta_{13}$ is significantly nonzero,
the two terms in Eq.~(\ref{mutau-IH}) can cancel each other since 
the $\theta_{13}$ term is enhanced by the factor $(1/\widehat{\rho})$. 
Thus to satisfy $\mu-\tau$ symmetry in $M_M$, either 
$\theta_{13}\approx 0 \approx \tilde\theta_{23}$, or 
the upper bound on $\theta_{13}$ should be large.
In the latter case, the actual value of $\sin \theta_{13}$ 
would have to be much smaller than 
$\widehat{\rho}$ to prevent the second term from becoming too large. 
These features can be seen from Fig.~\ref{IH_theta23}.
Indeed, for $\sin \theta_{13} \gsim 0.01$, it is possible for 
$M_M$ to satisfy the $\mu$-$\tau$ symmetry for any value of
$\tilde\theta_{23}$.
For smaller values of $\theta_{13}$, one needs 
$\tilde\theta_{23} \lsim 1.5^\circ$.

\subsection{$S_3$ symmetry in $M_M$}

The $S_3$ symmetry in $M_M$ implies, from (\ref{mM-def}), that
$xyB = xzC = yzE, x^2 A = y^2 D = z^2 F$. The Dirac-independent 
conditions that can be obtained are $(B/C)^2=(D/F)$ as in the
case of the $\mu-\tau$ exchange symmetry, and the additional condition
$(B/E)^2=(A/F)$ is required for $S_3$ symmetry to be satisfied in $M_M$.  
Therefore, the deviation from $S_3$ symmetry can be quantified as:  
\begin{equation} 
\widetilde\Delta_{S_3}=
\left|\frac{B^2}{C^2}-\frac{D}{F}\right|^2+\left|\frac{B^2}{E^2} 
-\frac{A}{F}\right|^2\,. 
\end{equation} 

In the NH scenario,
\beq
(B^2/E^2 - A/F)_{\rm NH} = 
{\cal O}(\tilde\rho,\tilde\epsilon,\theta_{13},\tilde\theta_{23})  \; ,
\eeq
so that, combined with Eq.~(\ref{mutau-NH}), 
it is conceivable that $S_3$ symmetry holds.

In the IH scenario, we get
\beq
(B^2/E^2 - A/F)_{\rm IH} = 
{\cal O}(\widehat\rho,\epsilon,\theta_{13},\tilde\theta_{23}) \;. \\
\eeq
The extent to which $S_3$ is satisfied then depends only upon
the extent to which the $\mu$--$\tau$ symmetry is obeyed.
The effect of the deviation from this symmetry on the mixing
angles is exactly the same as given in the $\mu$--$\tau$ case. 
The $S_3$ permutation symmetry in $M_M$ 
can thus be satisfied for either $\sin \theta_{13} \gsim 0.01$,
or $|\tilde{\theta}_{23}|\lsim 1.5^\circ$ 
and $\sin \theta_{13} \lsim 0.01$,
as can be seen in  Fig.~\ref{IH_theta23}.
In the latter case, the actual value of $\sin \theta_{13}$ needs to be
much smaller than $\widehat{\rho}$. 

For degenerate neutrino masses, 
\barr
(B^2/E^2 - A/F)_{\rm QD} & = & \frac{2 e^{2 i(\chi_e - \chi_\tau)} 
(e^{2 i \phi_1} - e^{2 i \phi_2})^2 c_{12}^2 s_{12}^2}{
(e^{2 i \phi_1} c_{12}^2 +  e^{2 i \phi_2} s_{12}^2
-  e^{2 i (\phi_1 + \phi_2)})^2} \nonumber \\
& & - \frac{ (e^{2 i \phi_2} c_{12}^2 -  e^{2 i \phi_1} s_{12}^2)}{  
e^{2 i \phi_1} c_{12}^2 +  e^{2 i \phi_2} s_{12}^2
-  e^{2 i (\phi_1 + \phi_2)}} 
\earr
Since at the moment we have no restrictions on the Majorana phases,
proper choice of the values of these phases would allow us to
make $(B^2/E^2 - A/F)_{\rm QD}$ vanish. Combining this with 
Eq.~(\ref{mutau-QD}), one can claim that it is possible to have
$\widetilde\Delta_{S_3} \approx 0$ for quasi-degenerate neutrinos.

\section{Conclusions and Outlook} 
\label{concl}

We examine texture zeroes and discrete flavor symmetries 
in the neutrino mass matrix through a bottom-up approach. 
We develop a formalism that uses the low energy data, checks 
its consistency with the desired texture or symmetry, and
quantifies the deviation from this symmetry.
To this end, we parameterize the neutrino mass matrix in terms 
of the masses, mixing angles, CP violating Dirac and Majorana phases,
as well as the ``flavor'' phases that have no
phenomenological implication but may be fixed by the mechanism of
neutrino mass generation.
We consider both the normal as well as inverted mass ordering of
neutrinos, and the cases when the neutrino masses are hierarchical
and quasidegenerate.
The results can be described analytically in the
three scenarios: normal ordering with hierarchical neutrino masses (NH),
inverted ordering with hierarchical neutrino masses (IH), and
any mass ordering, but quasidegenerate neutrinos (QD).
In each independent scenario, we identify parameters that are small
and hence can be used in a perturbative expansion.
This provides a universal framework to analytically understand many 
results that were only known numerically before.
In addition, it allows us to make new predictions about
symmetries that can be verified by numerical means.

In the bottom-up approach, the extent to which the symmetries can be
tested is limited by the accuracy of the experimental data on
neutrino parameters, and hence one can only talk about the symmetry
being approximately satisfied, or it being consistent with the
data. In order to achieve this, we quantify the deviation  from
such a symmetry in each case through a quantity that vanishes 
in the limit of exact symmetry. The minimum value of such a 
quantity that is consistent with the current data is an indication
of the extent to which the symmetry is valid.
If this minimum value is zero, the symmetry is clearly obeyed.
However, our formalism allows even for quantification of the breaking
of the symmetry in a model independent, unified way.
For illustration, in each scenario we consider the symmetry to be
viable if the relevant quantity is less than $10^{-2}$.

It is found that the viability of texture zeroes strongly depends
on the absolute neutrino mass scale, the nature of mass ordering,
and the angle $\theta_{13}$.
For the neutrino mass matrix $M_\nu$, all the six possible one-zero
textures are allowed, though only in certain scenarios.
For example, in normal mass ordering one can have only three
possible one-zero textures when the neutrino
masses are hierarchical, and five one-zero textures 
if they are quasidegenerate.
The inverted ordering allows for four one-zero textures 
in the hierarchical limit (IH) and five in the quasidegenerate
limit.
We find that seven two-zero textures are allowed in $M_\nu$,
while no three-zero texture is permitted.

The texture zeroes of the inverse mass matrix $M_\nu^{-1}$ correspond
to those of the Majorana mass matrix $M_M$ of the heavy neutrinos
in the Type-I seesaw mechanism, so we can treat them together.
We find that all six one-zero textures are allowed for $M_M$,
though only five of them are allowed for normal ordering.
For inverted ordering, only three are allowed if the neutrinos have
hierarchical masses, while five are allowed when the neutrinos
are quasidegenerate. 
Seven two-zero textures are found to be consistent with the current data, 
while no three-zero texture is permitted.

All the results with texture zeroes can be understood analytically 
in terms of the expansion in small parameters that we have introduced in
each scenario. In most of the scenarios, the allowed freedom in
choosing Dirac and Majorana phases helps us to take the mass matrix
closer to the exact symmetry. Wherever one cannot do that, there is
a special reason that can be understood in terms of our analytic framework.
Since the viability of the texture zeroes is very sensitive to 
the absolute mass scale $m_0$ and the mixing angle $\theta_{13}$, 
future experiments aimed at determining these quantities
will have a large impact, which can be gauged from our results.

We use the same technique to examine flavor symmetries like $\mu$--$\tau$ 
exchange and $S_3$. 
We define quantities involving elements of the $M_\nu$ and $M_M$
matrices that vanish identically for the exact symmetry.
In order for the symmetry to be viable, we require this quantity
to vanish to the zeroth order of the relevant small parameters.
 
In $M_\nu$, we find that the $\mu-\tau$ symmetry is viable in
IH and QD scenarios as long as the unphysical
phases are related by $|\chi_\mu - \chi_\tau| \approx \pi$.
In the NH scenario however, one needs 
$|\chi_\mu - \chi_\tau| \approx \pi$ and $\theta_{13} \ll
(1-\dmsq_{\rm atm}/ 4 m_0^2)$.
As far as the $S_3$ symmetry is concerned, the current data
allows it only if the neutrinos are quasidegenerate.

Since the Dirac masses of neutrinos in Type-I seesaw mechanism
are unknown, the constraints put by the discrete flavor symmetries
on $M_M$ are slightly weaker.
As a result, the $\mu-\tau$ exchange symmetry is viable for 
NH and QD, 
while in the IH scenario one needs 
$\theta_{13} \ll |1 + \dmsq_{\rm atm}/(4 m_0^2)|$.
The $S_3$ symmetry in $M_M$ also holds under these conditions.

This paper provides a universal formalism to analyze the symmetries
of neutrino mass matrices and illustrates its applications to 
a set of symmetries, viz. texture zeroes and discrete flavor
symmetries. 
Currently many of the texture zeroes and flavor symmetries
considered here seem to be viable.
However, improved data on the mixing angles and masses expected in
near future will be able to rule out some of these. 
Moreover, the formalism developed here can be applied 
to a much wider class of problems related to the structure of
neutrino mass matrix, that can give us more insights into the
mechanism of neutrino mass generation.

\section*{Acknowledgement}

NS was supported by the European Union through the Marie Curie Research 
and Training Network ``UniverseNet" (MRTN-CT-2006-035863). 
He also thanks to the hospitality at Tata Institute of Fundamental 
Research where a part of this work has been done. 
The work of AD was partly supported by the Max Planck - India
Partnergroup program between Max Planck Institute for Physics 
and Tata Institute of Fundamental Research.

 

\begin{thebibliography}{99}  
 
\bibitem{nu-fits}
  T.~Schwetz, M.~Tortola and J.~W.~F.~Valle,
  arXiv:0808.2016 [hep-ph];
%
  G.~L.~Fogli, E.~Lisi, A.~Marrone, A.~Palazzo and A.~M.~Rotunno,
  Phys.\ Rev.\ Lett.\  {\bf 101}, 141801 (2008)
  [arXiv:0806.2649 [hep-ph]];
%
  A.~Bandyopadhyay, S.~Choubey, S.~Goswami, S.~T.~Petcov and D.~P.~Roy,
  arXiv:0804.4857 [hep-ph].


\bibitem{pontecorvo}
 B.~Pontecorvo,
  Sov.\ Phys.\ JETP {\bf 7}, 172 (1958)
  [Zh.\ Eksp.\ Teor.\ Fiz.\  {\bf 34}, 247 (1957)];
%
  B.~Pontecorvo,
  Sov.\ Phys.\ JETP {\bf 26}, 984 (1968)
  [Zh.\ Eksp.\ Teor.\ Fiz.\  {\bf 53}, 1717 (1967)];
%
  V.~N.~Gribov and B.~Pontecorvo,
  Phys.\ Lett.\  B {\bf 28}, 493 (1969).


\bibitem{mns}
 Z.~Maki, M.~Nakagawa and S.~Sakata,
  Prog.\ Theor.\ Phys.\  {\bf 28}, 870 (1962).

\bibitem{pdg}
 C.~Amsler {\it et al.}  [Particle Data Group],
  Phys.\ Lett.\  B {\bf 667}, 1 (2008).

\bibitem{wmap} 
 S.~Hannestad,
  Phys.\ Rev.\ Lett.\  {\bf 95}, 221301 (2005)
  [arXiv:astro-ph/0505551].


\bibitem{one-zero} 
C.~Hagedorn, J.~Kersten and M.~Lindner,
  Phys.\ Lett.\  B {\bf 597}, 63 (2004)
[arXiv:hep-ph/0406103]\,;
%
A.~Merle and W.~Rodejohann,
  Phys.\ Rev.\  D {\bf 73} (2006) 073012 [arXiv:hep-ph/0603111].
%

\bibitem{texture-zero} P.~H.~Frampton, S.~L.~Glashow and D.~Marfatia,
  Phys.\ Lett.\  B {\bf 536}, 79 (2002) [arXiv:hep-ph/0201008];
%
Z.~Z.~Xing,
  Phys.\ Lett.\  B {\bf 530}, 159 (2002) [arXiv:hep-ph/0201151];
%
A.~Kageyama {\it et al.}, 
  Phys.\ Lett.\  B {\bf 538}, 96 (2002) [arXiv:hep-ph/0204291];
%
B.~R.~Desai, D.~P.~Roy and A.~R.~Vaucher,
  Mod.\ Phys.\ Lett.\  A {\bf 18}, 1355 (2003) [arXiv:hep-ph/0209035];
%
S.~Dev, S.~Kumar, S.~Verma and S.~Gupta,
  Phys.\ Rev.\  D {\bf 76}, 013002 (2007)
  [arXiv:hep-ph/0612102].
%

\bibitem{two-zero-textures} 
T.~Endoh, S.~Kaneko, S.~K.~Kang, T.~Morozumi and M.~Tanimoto,
  Phys.\ Rev.\ Lett.\  {\bf 89}, 231601 (2002)
  [arXiv:hep-ph/0209020];
%
K.~Bhattacharya {\it et al.}, 
  Phys.\ Rev.\  D {\bf 74}, 093001 (2006)
  [arXiv:hep-ph/0607272];
%
S.~Goswami and A.~Watanabe,
  arXiv:0807.3438 [hep-ph].
%

\bibitem{mu-tau} An incomplete list:
 T.~Fukuyama and H.~Nishiura,
  arXiv:hep-ph/9702253;
%
R.~N.~Mohapatra and S.~Nussinov,
  Phys.\ Rev.\  D {\bf 60}, 013002 (1999)
  [arXiv:hep-ph/9809415];
%
E.~Ma and M.~Raidal,
  Phys.\ Rev.\ Lett.\  {\bf 87}, 011802 (2001)
  [Erratum-ibid.\  {\bf 87}, 159901 (2001)]
  [arXiv:hep-ph/0102255];
%
 K.~R.~S.~Balaji, W.~Grimus and T.~Schwetz,
  Phys.\ Lett.\  B {\bf 508}, 301 (2001)
  [arXiv:hep-ph/0104035];
%
C.~S.~Lam,
  Phys.\ Lett.\  B {\bf 507}, 214 (2001)
  [arXiv:hep-ph/0104116];
%
W.~Grimus and L.~Lavoura,
  JHEP {\bf 0107}, 045 (2001)
  [arXiv:hep-ph/0105212];
%
W.~Grimus and L.~Lavoura,
  Acta Phys.\ Polon.\  B {\bf 32}, 3719 (2001)
  [arXiv:hep-ph/0110041];
%
 E.~Ma,
  Phys.\ Rev.\  D {\bf 66}, 117301 (2002)
  [arXiv:hep-ph/0207352];
%
 Y.~Koide {\it et al.}, 
  Phys.\ Rev.\  D {\bf 66}, 093006 (2002)
  [arXiv:hep-ph/0209333];
%
 P.~F.~Harrison and W.~G.~Scott,
  Phys.\ Lett.\  B {\bf 547}, 219 (2002)
  [arXiv:hep-ph/0210197];
%
A.~Ghosal,
  arXiv:hep-ph/0304090;
%
K.~Matsuda and H.~Nishiura,
  Phys.\ Rev.\  D {\bf 69}, 053005 (2004)
  [arXiv:hep-ph/0309272];
%
 W.~Grimus {\it et al.}, 
  Nucl.\ Phys.\  B {\bf 713}, 151 (2005)
  [arXiv:hep-ph/0408123];
%
S.~Choubey and W.~Rodejohann,
  Eur.\ Phys.\ J.\  C {\bf 40}, 259 (2005)
  [arXiv:hep-ph/0411190];
%
T.~Kitabayashi and M.~Yasue,
  Phys.\ Lett.\  B {\bf 621}, 133 (2005)
  [arXiv:hep-ph/0504212];
%
R.~N.~Mohapatra and W.~Rodejohann,
  Phys.\ Rev.\  D {\bf 72}, 053001 (2005)
  [arXiv:hep-ph/0507312];
%
 A.~Datta and P.~J.~O'Donnell,
  Phys.\ Rev.\  D {\bf 72}, 113002 (2005)
  [arXiv:hep-ph/0508314];
%
A.~S.~Joshipura,
  Eur.\ Phys.\ J.\  C {\bf 53}, 77 (2008)
  [arXiv:hep-ph/0512252].


\bibitem{s3}  An incomplete list:
  P.~F.~Harrison, D.~H.~Perkins and W.~G.~Scott,
  Phys.\ Lett.\  B {\bf 458}, 79 (1999)
  [arXiv:hep-ph/9904297];
%
P.~F.~Harrison, D.~H.~Perkins and W.~G.~Scott,
  Phys.\ Lett.\  B {\bf 530}, 167 (2002)
  [arXiv:hep-ph/0202074];
%
 Z.~Z.~Xing,
  Phys.\ Lett.\  B {\bf 533}, 85 (2002)
  [arXiv:hep-ph/0204049];
%
P.~F.~Harrison and W.~G.~Scott,
  Phys.\ Lett.\  B {\bf 557}, 76 (2003) [arXiv:hep-ph/0302025];
%
 W.~Grimus and L.~Lavoura,
  JHEP {\bf 0508}, 013 (2005)
  [arXiv:hep-ph/0504153];
%
W.~Grimus and L.~Lavoura,
  JHEP {\bf 0601}, 018 (2006)
  [arXiv:hep-ph/0509239];
%
Y.~Koide,
  Phys.\ Rev.\  D {\bf 73}, 057901 (2006)
[arXiv:hep-ph/0603069].
%

\bibitem{top-down} See for example R.~N.~Mohapatra and A.~Y.~Smirnov,
  Ann.\ Rev.\ Nucl.\ Part.\ Sci.\  {\bf 56}, 569 (2006)
  [arXiv:hep-ph/0603118]\,.


\bibitem{forbidden-textures} F.~Plentinger, G.~Seidl and W.~Winter,
  JHEP {\bf 0804}, 077 (2008)
  [arXiv:0802.1718 [hep-ph]].

\bibitem{seesaw}  P.~Minkowski,
  Phys.\ Lett.\  B {\bf 67}, 421 (1977);
M.~Gell-Mann, P.~Ramond and R.~Slansky
in {\it Supergravity} (P.~van Niewenhuizen and D.~Freedman, eds),
(Amsterdam), North Holland, 1979; T.~Yanagida in {\it Workshop
on Unified Theory and Baryon number in the Universe} (O. Sawada
and A.~Sugamoto, eds), (Japan), KEK 1979; R.N.~Mohapatra and
G.~Senjanovic, Phys.\ Rev.\ Lett. {\bf 44}, 912 (1980).  



\bibitem{leptogenesis} M.~Fukugita and T.~Yanagida, Phys. Lett.
{\bf B174}, 45 (1986); See for a recent review: 
W.~Buchmuller, R.~D.~Peccei and T.~Yanagida,
  Ann.\ Rev.\ Nucl.\ Part.\ Sci.\  {\bf 55}, 311 (2005)
  [arXiv:hep-ph/0502169].
 
\bibitem{ma-determinant} 
L.~Lavoura,
  Phys.\ Lett.\  B {\bf 609}, 317 (2005) [arXiv:hep-ph/0411232];
%
E.~Ma,
  Phys.\ Rev.\  D {\bf 71}, 111301 (2005) [arXiv:hep-ph/0501056].


\bibitem{dgr2} A.~Dighe, S.~Goswami and P.~Roy,
  Phys.\ Rev.\  D {\bf 73}, 071301 (2006)
  [arXiv:hep-ph/0602062].

\bibitem{sanjeevetal.07} The details of the parameter space can be found in 
S.~Dev and S.~Kumar,
  Mod.\ Phys.\ Lett.\  A {\bf 22}, 1401 (2007) [arXiv:hep-ph/0607048].

\end{thebibliography}
\end{document}